\documentclass[10pt,aps,prd,showpacs,showkeys,amssymb,preprintnumbers,nofootinbib,
superscriptaddress,notitlepage,eqsecnum]{revtex4}

\usepackage{amsmath}
\usepackage{graphicx}
\usepackage{latexsym}
\usepackage[usenames]{color}
\usepackage{amsfonts}
\usepackage{url,hyperref}
\usepackage{bm}

\hypersetup{colorlinks=true, citecolor=blue}
\definecolor{dred}{rgb}{0.75,0.08,0}

\newcommand{\nn}{\nonumber}
\newcommand{\beq}{\begin{equation}}
\newcommand{\eeq}{\end{equation}}
\def\bea{\begin{eqnarray}}
\def\eea{\end{eqnarray}}

\begin{document}

\title{Quasi-normal modes for doubly rotating black holes}
\author{H.~T.~Cho}
\email[Email: ]{htcho@mail.tku.edu.tw}
\affiliation{Department of Physics, Tamkang University, Tamsui, Taipei, Taiwan, Republic of 
China}
\author{Jason~Doukas}
\email[Email: ]{jasonad@yukawa.kyoto-u.ac.jp}
\affiliation{Yukawa Institute for Theoretical Physics, Kyoto University, Kyoto, 606-8502, Japan}

\author{Wade~Naylor}
\email[Email: ]{naylor@het.phys.sci.osaka-u.ac.jp}
\affiliation{International College \& Department of Physics, Osaka University, Toyonaka, 
 Osaka 560-0043, Japan}

\author{A.~S.~Cornell}
\email[Email: ]{alan.cornell@wits.ac.za}
\affiliation{National Institute for Theoretical Physics; School of Physics, University of the Witwatersrand, Wits 2050, South Africa}

\begin{abstract}
Based on the work of Chen, L\"u and Pope, we derive expressions for the $D\geq 6$  dimensional metric for Kerr-(A)dS black holes with two independent rotation parameters and all others set equal to zero:  $a_1\neq 0, a_2\neq0, a_{3}=a_4=\cdots=0$. The Klein-Gordon equation is then explicitly separated on this background.  For $D\geq 6$ this separation results in a radial equation coupled to two generalized spheroidal angular equations. We then develop a full numerical approach that utilizes the Asymptotic Iteration Method (AIM) to find radial Quasi-Normal Modes (QNMs) of doubly rotating flat Myers-Perry black holes for slow rotations. We also develop perturbative expansions for the angular quantum numbers in powers of the rotation parameters up to second order.
\end{abstract}

\date{\today}
\preprint{YITP-11-44, WITS-CTP-68, OU-HET-701/2011}
\maketitle

\section{Introduction}\label{sec:intro}

\par After the advent of the brane world scenario \cite{Randall:1999ee} and the AdS/CFT correspondence \cite{Maldacena:1997re}, there has been a growing interest in the study of higher-dimensional black holes. Perhaps the strongest driver behind this interest is that the Large Hadron Collider (LHC) \cite{Kanti:2004nr} may produce black holes in the near future and such black holes, if produced, would have large angular momentum. Therefore higher dimensional generalisations of the Kerr solution are the natural setting for these studies. 

\par Rotating black holes in higher dimensions were first discussed in the seminal paper by Myers and Perry \cite{Myers:1986un}.  One of the unexpected results to come from this work was that some families of solutions were shown to have event horizons for arbitrarily large values of their rotation parameters. The stability of such black holes is certainly in question \cite{EmparanMyers, Konoplya:2011qq}, but no direct proof of instability has been provided. Another new feature of the Myers-Perry (MP) solutions was that they in general have $\lfloor\frac{D-1}{2}\rfloor$
 spin parameters, making them somewhat more complex than the four dimensional Kerr solution. The first asymptotically non-flat five-dimensional MP metric was given in \cite{Hawking:1998kw}. Subsequent generalizations to arbitrary dimensions was done in \cite{Gibbons:2004js}, and finally the most general Kerr-(A)dS-NUT metric was found by Chen, L\"u and Pope \cite{Chen:2006xh}.

\par In the literature the focus has been largely directed toward solutions with only one rotation parameter, the so called simply-rotating case. The reason for this is that in the brane world, collider produced black holes would initially only have one dominant angular momentum direction. This is due to the particles that produce the black-hole being confined to our 3-brane and therefore the rotation would be largest in the plane of collision on the brane.  However, there is reason to believe this picture may be too naive. In any realistic situation the brane would be expected to have a thickness of the inverse Plank scale. At impact, the colliding particles would in general be offset in these thick directions and therefore further non-zero angular momenta would be present in other rotation planes of the black hole. Even though these angular momenta would be small compared with the rotation on the brane there is strong evidence \cite{Nomura:2005mw} to suggest that such black holes would evolve into a final state in which all the angular momentum parameters were of the same order. There are also compelling theoretical reasons why one would want to go beyond the simply-rotating case. In particular, the Quasi-Normal Modes (QNMs) of these solutions may have applications in the AdS/CFT correspondence.  

\par The study of the wave equations in higher dimensional rotating black hole spacetimes was initiated in \cite{Frolov:2002xf} by analyzing the Klein-Gordon equation in five dimensions. The analysis relied crucially on the method of the separation of variables. The problem of separability of these wave equations in higher dimensions is a difficult one, even for the Klein-Gordon case; early attempts were only aimed at special cases \cite{Vasudevan:2005js}. Finally, using the Chen-L\"u-Pope metric, Frolov, Krtous and Kubiznak \cite{Frolov:2006pe} were able to separate the geodesic equation and the Klein-Gordon equation in the most general setting. This was then realized to be due to the presence of hidden symmetries, in the form of Killing tensors \cite{Krtous:2006qy, Frolov:2007nt, Frolov:2008jr}. A whole tower of Killing tensors and symmetry operators \cite{Sergyeyev:2007gf} can be constructed with the help of the corresponding Killing-Yano and conformal Killing-Yano tensors. They guarantee the separability of the geodesic equation, the Klein-Gordon equation and also the Dirac equation \cite{Oota:2007vx}. Unfortunately, the higher spin wave equations, especially the gravitational perturbation equation, have not yet been subjected to such an analysis.

\par Even for the scalar wave equation, efforts so far have been focused mostly on the simply-rotating case. Notably, in \cite{Morisawa:2004fs, Cardoso:2004cj, Kodama:2009bf}, the stability of the scalar perturbation in six and higher dimensions was considered in the ultra-rotating cases, where no instability was found. Due to the interest in AdS spacetimes, the investigation was extended to Kerr-AdS black holes \cite{Kodama:2009rq, Uchikata:2009zz,Cardoso:2006wa, Cardoso:2004nk, Cardoso:2004hs}. Here the expected superradiant instability did indeed show up. In addition, Hawking radiation in these spacetimes (Kerr-dS) \cite{Doukas:2009cx,Kanti:2009sn} were calculated with possible applications to the production and decay of LHC black holes \cite{Kanti:2009sz}.

\par In this article we show how both the $D\geq 6$ two-rotation metric and the separation of the scalar wave-equation on this metric can be achieved quite economically by working with the general Kerr-NUT-AdS metric described in \cite{Chen:2006xh}. That is, we begin with the full set of $\lfloor \frac{D-1}{2} \rfloor$ rotation parameters, $a_\alpha$, and $\lfloor \frac{D-2}{2} \rfloor$ coordinate variables, $y_{i}$, and then take the limit that all but two of the rotations goes to zero. This then reduces the full metric down to that with only two non-zero rotation parameters and allows us to present this metric explicitly in these coordinates, with an expression valid for any dimension $D\geq 6$ \footnote{The case with $D=5$ is exceptional because even though there are two rotation parameters there is only one Jacobi coordinate variable $y_1$. As we shall see, for $D\geq 6$, in the limit of keeping only two rotation parameters a total of two Jacobi coordinates survive. The $D=5$ case will be considered in a separate work \cite{Cho:inPROG}.}.

\par In this form the separation of the Klein-Gordon equation proceeds analogously to the case with all rotations present. For simplicity we will set the NUT charges $L_\alpha =0$ although in general this is not an obstacle to separation. 

\par The structure of the paper is as follows: In the next section (Section \ref{sec:KG}), we present the general metric of Kerr-(A)dS black holes with two rotations. The corresponding Klein-Gordon equation is separated into one radial equation and two angular ones. In section \ref{numeric} we develop a full numerical method using the Asymptotic Iteration Method (AIM) to solve for the angular eigenvalues and the radial quasinormal modes (QNMs) for two rotation parameters. Conclusions and discussions are then given in Section \ref{conc}. 
In Appendix \ref{Ang6} the small $a_1,a_2$ expansion for the angular quantum numbers for two angular equations analytically are given up to second order; we call this ``double perturbation theory". This allows us to check the AIM against the perturbative method.

%
%

\section{The metric and the separated equations of the Klein-Gordon equation with two rotations}
\label{sec:KG}
In this section, we shall first derive the Kerr-(A)dS metric with two rotations from the general metric obtained in \cite{Chen:2006xh}. To specialize to the case with only two rotations, we take the limit $a_{3}$, $a_{4},\cdots\rightarrow 0$, while, without loss of generality, keep $a_{1}>a_{2}>a_{3}>a_{4}>\cdots$, where the $a_{i}$'s are the rotation parameters. Then we explicitly separate the Klein-Gordon equation on this metric. The separation in the general Kerr-(A)dS metric was first performed in \cite{Frolov:2006pe}. We follow their procedure for the case with only two rotations, where we find one resulting radial equation and two angular ones.

\subsection{General metric with two rotations}

We start looking at the metric with $D=2n$, that is, for even dimensions. However, the result we obtain at the end is also valid for odd dimensions. For $D=2n$, there are at most $n-1$ rotation directions so we have $a_{i}$, with $i=1,2,\dots,n-1$. This metric, satisfying $R_{\mu\nu}=-3g^{2}g_{\mu\nu}$, can be expressed as follows \cite{Chen:2006xh}.
\begin{eqnarray}
ds^{2}&=&\frac{U}{X}dr^{2}+\sum_{\alpha=1}^{n-1}\frac{U_{\alpha}}{X_{\alpha}}dy_{\alpha}^{2}
-\frac{X}{U}\left[Wd\tilde{t}-\sum_{i=1}^{n-1}\gamma_{i}d\tilde{\phi}_{i}\right]^{2}
+\ \ \sum_{\alpha=1}^{n-1}\frac{X_{\alpha}}{U_{\alpha}}\left[\frac{(1+g^{2}r^{2})W}{1-g^{2}y_{\alpha}^{2}}d\tilde{t}
-\sum_{i=1}^{n-1}\frac{(r^{2}+a_{i}^{2})\gamma_{i}}{a_{i}^{2}-y_{\alpha}^{2}}d\tilde{\phi}_{i}\right]^{2},\label{metric}
\end{eqnarray}
where
\begin{eqnarray}
U&=&\prod_{\alpha=1}^{n-1}(r^{2}+y_{\alpha}^{2})\;,\ \ \ \ \ \ U_{\alpha}=-(r^{2}+y_{\alpha}^{2})\prod_{\beta=1,\beta\neq\alpha}^{n-1}(y_{\beta}^{2}-y_{\alpha}^{2})\;,\\
X&=&(1+g^{2}r^{2})\prod_{k=1}^{n-1}(r^{2}+a_{k}^{2})-2Mr\;,\ \ \ \ \ \ X_{\alpha}=-(1-g^{2}y_{\alpha}^{2})\prod_{k=1}^{n-1}(a_{k}^{2}-y_{\alpha}^{2})\;,\\
W&=&\prod_{\alpha=1}^{n-1}(1-g^{2}y_{\alpha}^{2})\;,\ \ \ \ \ \ \gamma_{i}=\prod_{\alpha=1}^{n-1}(a_{i}^{2}-y_{\alpha}^{2})\;,\\
t&=&\tilde{t}\prod_{i=1}^{n-1}(1-g^{2}a_{i}^{2})\;,\ \ \ \ \ \ \phi_{i}=a_{i}(1-g^{2}a_{i}^{2})\tilde{\phi}_{i}\prod_{k=1,k\neq i}^{n-1}(a_{i}^{2}-a_{k}^{2})\;,
\end{eqnarray}
with $1\leq\alpha,i\leq n-1$. $\phi_{i}$ is the azimuthal angle for each $a_{i}$\footnote{Note that the metric for the odd case is slightly different, see \cite{Chen:2006xh}. Chen {\it{et al.}} also define an extra parameter $a_n=0$ in the even case in order to make some parts of the treatment between even and odd cases homogeneous. However, for better clarity we have elected not to do this here, i.e., we assume there are only $n-1$ parameters, $a_i$, in the even case.}. With the direction cosines $\mu_{i}$, $i=1,\dots,n$, the metric for a unit $D-2$ sphere is just
\beq
d\Omega^{2}=\sum_{i=1}^{n}d\mu_{i}^{2}+\sum_{i=1}^{n-1}\mu_{i}^{2}d\phi_{i}^{2}\;,
\eeq
subject to the constraint $\sum_{i=1}^{n}\mu_{i}^{2}=1$. This constraint can be solved in terms of the unconstrained latitude variables $y_{\alpha}$'s,
\begin{equation}
\mu_{i}^{2}=\frac{\prod_{\alpha=1}^{n-1}(a_{i}^{2}-y_{\alpha}^{2})}{a_{i}^{2}\prod_{k=1,k\neq i}^{n-1}(a_{i}^{2}-a_{k}^{2})}\;,\ \ \ \ \ \mu_{n}^{2}=\frac{\prod_{\alpha=1}^{n-1}y_{\alpha}^{2}}{\prod_{i=1}^{n-1}a_{i}^{2}}\;.\label{defy}
\end{equation}
Expressed in terms of the unconstrained $y_i$ coordinates the metric for the unit sphere becomes diagonal:  
\begin{equation}
d\Omega^{2}=\sum_{\alpha=1}^{n-1}g_{\alpha}dy_{\alpha}^{2}+\sum_{i=1}^{n-1}\mu_{i}^{2}d\phi_{i}^{2}\;,\label{spheremetric}
\end{equation}
with
\begin{equation}
g_{\alpha}=\frac{\prod_{\beta=1,\beta\neq\alpha}^{n-1}(y_{\beta}^{2}-y_{\alpha}^{2})}{\prod_{k=1}^{n-1}(a_{k}^{2}-y_{\alpha}^{2})}\;.
\end{equation}
This choice then allows for a more symmetric form of the general Kerr-(A)dS metrics \cite{Chen:2006xh} as we shall see below for the two rotations case.

\par To obtain a general metric with two rotations we take the limit $a_{3}$, $a_{4}$, $\dots$, $a_{n-1}\rightarrow 0$ while assuming that $a_{1}>a_{2}>a_{3}>\cdots>a_{n-1}$. From the definition in Eq.~(\ref{defy}) we see that $y_{i}$ is of the same order of magnitude as $a_{i}$. 

\par Bearing this in mind, in the above limit, we can consider the different terms in equation (\ref{metric}):
\begin{eqnarray}
\frac{U}{X}dr^{2}&=&\frac{(r^{2}+y_{1}^{2})(r^{2}+y_{2}^{2})}{\Delta_{r}}dr^{2} \, , 
\end{eqnarray}
where 
\beq
\Delta_{r}=(1+g^{2}r^{2})(r^{2}+a_{1}^{2})(r^{2}+a_{2}^{2})-{2M\over r^{D-7}} \, .
\eeq 
Note the solutions of $\Delta_r=0$ lead to the black hole and cosmological horizons: $r_+$ and $r_c$, respectively, for the Kerr-dS case ($g<0$). 

\par As another example, under this limiting procedure, the quantities
\begin{equation}
U_{3}\rightarrow -r^{2}y_{1}^{2}y_{2}^{2}(-y_{3}^{2})^{n-4},\ \ \ X_{3}\rightarrow -a_{1}^{2}a_{2}^{2}(a_{3}^{2}-y_{3}^{2})(-y_{3}^{2})^{n-4}\,,
\end{equation}
both seem to vanish in the limiting process, however, the ratio
\begin{equation}
\frac{U_{3}}{X_{3}}dy_{3}^{2}\rightarrow r^{2}\left(\frac{y_{1}^{2}y_{2}^{2}}{a_{1}^{2}a_{2}^{2}}\right)\frac{1}{a_{3}^{2}-y_{3}^{2}}dy_{3}^{2}
=r^{2}\left(\frac{y_{1}^{2}y_{2}^{2}}{a_{1}^{2}a_{2}^{2}}\right)g_{3}dy_{3}^{2}
\end{equation}
is actually finite. Here in defining $g_{3}$ we have taken into account only the angular variables associated with $a_{3}$, $a_{4}$, $\dots$, $a_{n-1}$. In the same way, part of the sum in the second term of the metric in Eq.~(\ref{metric}) then constitutes
\begin{equation}
\sum_{\alpha=3}^{n-1}\frac{U_{\alpha}}{X_{\alpha}}dy_{\alpha}^{2}\rightarrow r^{2}\left(\frac{y_{1}^{2}y_{2}^{2}}{a_{1}^{2}a_{2}^{2}}\right)\sum_{\alpha=3}^{n-1}g_{\alpha}dy_{\alpha}^{2}\;.
\end{equation}
Similarly, part of the other sum in Eq.~(\ref{metric}) gives
\begin{equation}
\sum_{\alpha=3}^{n-1}\frac{X_{\alpha}}{U_{\alpha}}\left[\frac{(1+g^{2}r^{2})W}{1-g^{2}y_{\alpha}^{2}}d\tilde{t}
-\sum_{i=1}^{n-1}\frac{(r^{2}+a_{i}^{2})\gamma_{i}}{a_{i}^{2}-y_{\alpha}^{2}}d\tilde{\phi}_{i}\right]^{2}\rightarrow
r^{2}\left(\frac{y_{1}^{2}y_{2}^{2}}{a_{1}^{2}a_{2}^{2}}\right)\sum_{i=3}^{n-1}\mu_{i}^{2}d\phi_{i}^{2}\;.
\end{equation}
Combining these two we obtain the metric for a $D-6$ sphere,
\begin{equation}
r^{2}\left(\frac{y_{1}^{2}y_{2}^{2}}{a_{1}^{2}a_{2}^{2}}\right)\left(\sum_{\alpha=3}^{n-1}g_{\alpha}dy_{\alpha}^{2}+
\sum_{i=3}^{n-1}\mu_{i}^{2}d\phi_{i}^{2}\right)
=r^{2}\left(\frac{y_{1}^{2}y_{2}^{2}}{a_{1}^{2}a_{2}^{2}}\right)d\Omega_{D-6}^{2}\;,
\end{equation}
as indicated in Eq.~(\ref{spheremetric}).

Finally, for dimensions $D\geq 6$, the metric with two rotations can be given by
\begin{eqnarray}
ds^{2}&=&-\frac{\Delta_{r}}{(r^{2}+y_{1}^{2})(r^{2}+y_{2}^{2})}
\left[\frac{(1-g^{2}y_{1}^{2})(1-g^{2}y_{2}^{2})}{(1-g^{2}a_{1}^{2})(1-g^{2}a_{2}^{2})}dt-\frac{(a_{1}^{2}-y_{1}^{2})(a_{1}^{2}-y_{2}^{2})}
{(1-g^{2}a_{1}^{2})(a_{1}^{2}-a_{2}^{2})}\frac{d\phi_{1}}{a_{1}}\right.\nonumber\\
&&\ \ \ \ \ \ \ \ \ \ \ \ \ \ \ \ \ \ \ \ \ \ \ \ \ \ \ \ \ \ \ \ \ \ \ \ \ \left.-\frac{(a_{2}^{2}-y_{1}^{2})(a_{2}^{2}-y_{2}^{2})}
{(1-g^{2}a_{2}^{2})(a_{2}^{2}-a_{1}^{2})}\frac{d\phi_{2}}{a_{2}}\right]^{2}\nonumber\\
&&+\frac{\Delta_{y_{1}}}{(r^{2}+y_{1}^{2})(y_{2}^{2}-y_{1}^{2})}
\left[\frac{(1+g^{2}r^{2})(1-g^{2}y_{2}^{2})}{(1-g^{2}a_{1}^{2})(1-g^{2}a_{2}^{2})}dt-\frac{(r^{2}+a_{1}^{2})(a_{1}^{2}-y_{2}^{2})}
{(1-g^{2}a_{1}^{2})(a_{1}^{2}-a_{2}^{2})}\frac{d\phi_{1}}{a_{1}}\right.\nonumber\\
&&\ \ \ \ \ \ \ \ \ \ \ \ \ \ \ \ \ \ \ \ \ \ \ \ \ \ \ \ \ \ \ \ \ \ \ \ \ \left.-\frac{(r^{2}+a_{2}^{2})(a_{2}^{2}-y_{2}^{2})}
{(1-g^{2}a_{2}^{2})(a_{2}^{2}-a_{1}^{2})}\frac{d\phi_{2}}{a_{2}}\right]^{2}\nonumber\\
&&+\frac{\Delta_{y_{2}}}{(r^{2}+y_{2}^{2})(y_{1}^{2}-y_{2}^{2})}
\left[\frac{(1+g^{2}r^{2})(1-g^{2}y_{1}^{2})}{(1-g^{2}a_{1}^{2})(1-g^{2}a_{2}^{2})}dt-\frac{(r^{2}+a_{1}^{2})(a_{1}^{2}-y_{1}^{2})}
{(1-g^{2}a_{1}^{2})(a_{1}^{2}-a_{2}^{2})}\frac{d\phi_{1}}{a_{1}}\right.\nonumber\\
&&\ \ \ \ \ \ \ \ \ \ \ \ \ \ \ \ \ \ \ \ \ \ \ \ \ \ \ \ \ \ \ \ \ \ \ \ \ \left.-\frac{(r^{2}+a_{2}^{2})(a_{2}^{2}-y_{1}^{2})}
{(1-g^{2}a_{2}^{2})(a_{2}^{2}-a_{1}^{2})}\frac{d\phi_{2}}{a_{2}}\right]^{2}\nonumber\\
&&+\frac{(r^{2}+y_{1}^{2})(r^{2}+y_{2}^{2})}{\Delta_{r}}dr^{2}+\frac{(r^{2}+y_{1}^{2})(y_{2}^{2}-y_{1}^{2})}{\Delta_{y_{1}}}dy_{1}^{2}
+\frac{(r^{2}+y_{2}^{2})(y_{1}^{2}-y_{2}^{2})}{\Delta_{y_{2}}}dy_{2}^{2}\nonumber\\
&&+r^{2}\left(\frac{y_{1}^{2}y_{2}^{2}}{a_{1}^{2}a_{2}^{2}}\right)d\Omega^{2}_{D-6}\; ,
\label{kerradsmetric}
\end{eqnarray}
and
\begin{eqnarray}
\Delta_{r}&=&(1+g^{2}r^{2})(r^{2}+a_{1}^{2})(r^{2}+a_{2}^{2})-2Mr^{7-D},\\
\Delta_{y_{1}}&=&(1-g^{2}y_{1}^{2})(a_{1}^{2}-y_{1}^{2})(a_{2}^{2}-y_{1}^{2})\;,\\
\Delta_{y_{2}}&=&(1-g^{2}y_{2}^{2})(a_{1}^{2}-y_{2}^{2})(a_{2}^{2}-y_{2}^{2})\;.
\end{eqnarray}
Here we have kept the variables $y_{1}$ and $y_{2}$ instead of writing them in terms of angular variables. This is because the relationship, as shown in Eq.~(\ref{defy}), is rather complicated to write out explicitly. If we solve $y_{1}$ and $y_{2}$ in terms of $\mu_{1}$ and $\mu_{2}$, we have
\begin{eqnarray}
y_{1,2}^{2}=\frac{1}{2}\left[a_{1}^{2}(1-\mu_{1}^{2})+a_{2}^{2}(1-\mu_{2}^{2})\pm\sqrt{4a_{1}^{2}a_{2}^{2}(\mu_{1}^{2}+\mu_{2}^{2}-1)
+(a_{1}^{2}(1-\mu_{1}^{2})+a_{2}^{2}(1-\mu_{2}^{2}))^{2}}\right].
\end{eqnarray}
It then follows that $y_{1}$ and $y_{2}$ must be constrained by
\begin{eqnarray}\label{yconstraints}
  a_{2}\leq y_{1}\leq a_{1}\ \ \ \ \ ;\ \ \ \ \ 0\leq y_{2}\leq a_{2}
\end{eqnarray}
in order for Eq.~(\ref{defy}) to be well-defined.

\par We used a similar procedure as above to show that the metric in the odd dimensional $D=2n+1$ case reduces to the same form as the one shown in equation (\ref{kerradsmetric}).

\subsection{Separated equations for the Klein-Gordon equation}
\label{separat}

The separation of the Klein-Gordon equation in the general Kerr-(A)dS metric has been achieved in \cite{Frolov:2006pe}. Here we shall show explicitly how the separation goes for the case with two rotations. To begin with it is convenient to rewrite the metric in Eq.~(\ref{kerradsmetric}) as
\begin{eqnarray}
ds^{2}&=&-Q_{1}\left[A_{1}^{(0)}d\psi_{0}+A_{1}^{(1)}d\psi_{1}+A_{1}^{(2)}d\psi_{2}\right]^{2}
+Q_{2}\left[A_{2}^{(0)}d\psi_{0}+A_{2}^{(1)}d\psi_{1}+A_{2}^{(2)}d\psi_{2}\right]^{2}\nonumber\\
&&\ \ \ \ +Q_{3}\left[A_{3}^{(0)}d\psi_{0}+A_{3}^{(1)}d\psi_{1}+A_{3}^{(2)}d\psi_{2}\right]^{2}+\frac{1}{Q_{1}}dr^{2}+\frac{1}{Q_{2}}dy_{1}^{2}
+\frac{1}{Q_{3}}dy_{2}^{2}+r^{2}\left(\frac{y_{1}^{2}y_{2}^{2}}{a_{1}^{2}a_{2}^{2}}\right)d\Omega_{D-6}^{2}\;,
\end{eqnarray}
where
\begin{eqnarray}
Q_{1}=\frac{\Delta_{r}}{(r^{2}+y_{1}^{2})(r^{2}+y_{2}^{2})}\ \ \ ;\ \ \
Q_{2}=\frac{\Delta_{y_{1}}}{(r^{2}+y_{1}^{2})(y_{2}^{2}-y_{1}^{2})}\ \ \ ;\ \ \
Q_{3}=\frac{\Delta_{y_{2}}}{(r^{2}+y_{2}^{2})(y_{1}^{2}-y_{2}^{2})}\;,
\end{eqnarray}
and $\psi_{k}$ are related to $t$, $\phi_{1}$ and $\phi_{2}$ by
\begin{eqnarray}
\psi_{0}&=&\frac{t}{(1-g^{2}a_{1}^{2})(1-g^{2}a_{2}^{2})}-\frac{a_{1}^{3}\phi_{1}}{(1-g^{2}a_{1}^{2})(a_{1}^{2}-a_{2}^{2})}
-\frac{a_{2}^{3}\phi_{2}}{(1-g^{2}a_{2}^{2})(a_{2}^{2}-a_{1}^{2})}\\
\psi_{1}&=&-\frac{g^{2}t}{(1-g^{2}a_{1}^{2})(1-g^{2}a_{2}^{2})}+\frac{a_{1}\phi_{1}}{(1-g^{2}a_{1}^{2})(a_{1}^{2}-a_{2}^{2})}
+\frac{a_{2}\phi_{2}}{(1-g^{2}a_{2}^{2})(a_{2}^{2}-a_{1}^{2})}\\
\psi_{2}&=&\frac{g^{4}t}{(1-g^{2}a_{1}^{2})(1-g^{2}a_{2}^{2})}-\frac{\phi_{1}}{a_{1}(1-g^{2}a_{1}^{2})(a_{1}^{2}-a_{2}^{2})}
-\frac{\phi_{2}}{a_{2}(1-g^{2}a_{2}^{2})(a_{2}^{2}-a_{1}^{2})}\;,
\end{eqnarray}
or conversely,
\begin{eqnarray}
t&=&\psi_{0}+(a_{1}^{2}+a_{2}^{2})\psi_{1}+a_{1}^{2}a_{2}^{2}\psi_{2}\label{tpsi}\\
\frac{\phi_{1}}{a_{1}}&=&g^{2}\psi_{0}+(1+g^{2}a_{2}^{2})\psi_{1}+a_{2}^{2}\psi_{2}\label{phi1psi}\\
\frac{\phi_{2}}{a_{2}}&=&g^{2}\psi_{0}+(1+g^{2}a_{1}^{2})\psi_{1}+a_{1}^{2}\psi_{2}\;.\label{phi2psi}
\end{eqnarray}
The matrix $A_{\mu}^{(k)}$ is given by
\begin{eqnarray}
A_{\mu}^{(k)}=\left(
\begin{array}{ccc}
1 & y_{1}^{2}+y_{2}^{2} & y_{1}^{2}y_{2}^{2} \\
1 & -r^{2}+y_{2}^{2} & -r^{2}y_{2}^{2} \\
1 & -r^{2}+y_{1}^{2} & -r^{2}y_{1}^{2}
\end{array}
\right),
\end{eqnarray}
with the inverse $B_{(k)}^{\mu}$,
\begin{eqnarray}
B_{(k)}^{\mu}=\left(
\begin{array}{ccc}
\frac{r^{4}}{(r^{2}+y_{1}^{2})(r^{2}+y_{2}^{2})} & \frac{-y_{1}^{4}}{(r^{2}+y_{1}^{2})(y_{2}^{2}-y_{1}^{2})} & \frac{-y_{2}^{4}}{(r^{2}+y_{2}^{2})(y_{1}^{2}-y_{2}^{2})} \\
\frac{r^{2}}{(r^{2}+y_{1}^{2})(r^{2}+y_{2}^{2})} & \frac{y_{1}^{2}}{(r^{2}+y_{1}^{2})(y_{2}^{2}-y_{1}^{2})} & \frac{y_{2}^{2}}{(r^{2}+y_{2}^{2})(y_{1}^{2}-y_{2}^{2})} \\
\frac{1}{(r^{2}+y_{1}^{2})(r^{2}+y_{2}^{2})} & \frac{-1}{(r^{2}+y_{1}^{2})(y_{2}^{2}-y_{1}^{2})} &
\frac{-1}{(r^{2}+y_{2}^{2})(y_{1}^{2}-y_{2}^{2})} \\
\end{array}
\right).
\end{eqnarray}
In this notation the inverse metric components are
\begin{eqnarray}
g^{rr}=Q_{1}\ \ \ ;\ \ \ g^{y_{1}y_{1}}=Q_{2}\ \ \ ;\ \ \ g^{y_{2}y_{2}}=Q_{3}\ \ \ ;\ \ \
g^{\psi_{i}\psi_{j}}=-\frac{1}{Q_{1}}B^{1}_{(i)}B^{1}_{(j)}+\frac{1}{Q_{2}}B^{2}_{(i)}B^{2}_{(j)}+\frac{1}{Q_{3}}B^{3}_{(i)}B^{3}_{(j)}\;,
\end{eqnarray}
plus the angular part related to the metric $g_{ab}$, and the inverse $g^{ab}$, for a unit $(D-6)$-dimensional sphere $S^{D-6}$. The determinant of the metric is then given by
\begin{eqnarray}
{\rm det}g_{\mu\nu}=-\left(r^{2}\frac{y_{1}^{2}y_{2}^{2}}{a_{1}^{2}a_{2}^{2}}\right)^{D-6}(r^{2}+y_{1}^{2})^{2}(r^{2}+y_{2}^{2})^{2}(y_{1}^{2}
-y_{2}^{2})^{2}{\rm det}g_{ab}\;.
\end{eqnarray}

Writing the Klein-Gordon field as
\begin{eqnarray}\label{eqn:SeparationAnsatz}
\Phi=R_{r}(r)R_{y_{1}}(y_{1})R_{y_{2}}(y_{2})e^{i\psi_{0}\Psi_{0}}e^{i\psi_{1}\Psi_{1}}e^{i\psi_{2}\Psi_{2}}Y(\Omega)\;,
\end{eqnarray}
the Klein-Gordon equation $\partial_{\mu}\left(\sqrt{-g}g^{\mu\nu}\partial_{\nu}\Phi\right)=0$ can be simplified to
\begin{eqnarray}
&&\frac{1}{(r^{2}+y_{1}^{2})(r^{2}+y_{2}^{2})}
\left[\frac{1}{R_{r}r^{D-6}}\partial_{r}\left(r^{D-6}\Delta_{r}\partial_{r}R_{r}\right)\right]\nonumber\\
&&+\frac{1}{(r^{2}+y_{1}^{2})(y_{2}^{2}-y_{1}^{2})}\left\{\frac{1}{R_{y_{1}}}\left(\frac{a_{1}}{y_{1}}\right)^{D-6}\partial_{y_{1}}
\left[\left(\frac{y_{1}}{a_{1}}\right)^{D-6}(1-g^{2}y_{1}^{2})(a_{1}^{2}-y_{1}^{2})(a_{2}^{2}-y_{1}^{2})
\partial_{y_{1}}R_{y_{1}}\right]\right\}\nonumber\\
&&+\frac{1}{(r^{2}+y_{2}^{2})(y_{1}^{2}-y_{2}^{2})}\left\{\frac{1}{R_{y_{2}}}\left(\frac{a_{2}}{y_{2}}\right)^{D-6}\partial_{y_{2}}
\left[\left(\frac{y_{2}}{a_{2}}\right)^{D-6}(1-g^{2}y_{2}^{2})(a_{1}^{2}-y_{2}^{2})(a_{2}^{2}-y_{2}^{2})
\partial_{y_{2}}R_{y_{2}}\right]\right\}\nonumber\\
&&+\frac{(r^{2}+y_{1}^{2})(r^{2}+y_{2}^{2})}{\Delta_{r}}\left[B_{(0)}^{1}\Psi_{0}+B_{(1)}^{1}\Psi_{1}+B_{(2)}^{1}\Psi_{2}\right]^{2}
-\frac{(r^{2}+y_{1}^{2})(y_{2}^{2}-y_{1}^{2})}{\Delta_{y_{1}}}
\left[B_{(0)}^{2}\Psi_{0}+B_{(1)}^{2}\Psi_{1}+B_{(2)}^{2}\Psi_{2}\right]^{2}\nonumber\\
&&-\frac{(r^{2}+y_{2}^{2})(y_{1}^{2}-y_{2}^{2})}{\Delta_{y_{2}}}
\left[B_{(0)}^{3}\Psi_{0}+B_{(1)}^{3}\Psi_{1}+B_{(2)}^{3}\Psi_{2}\right]^{2}-\frac{a_{1}^{2}a_{2}^{2}}{r^{2}y_{1}^{2}y_{2}^{2}}j(j+D-7)=0\;,
\end{eqnarray}
where $-j(j+D-7)$ is the eigenvalue of the Laplacian on $S^{D-6}$. By putting in the values of $B_{(k)}^{\mu}$ and by using the identities
\begin{eqnarray}
&&\frac{1}{r^{2}y_{1}^{2}y_{2}^{2}}=\frac{1}{(r^{2}+y_{1}^{2})(r^{2}+y_{2}^{2})r^{2}}+\frac{1}{(r^{2}+y_{1}^{2})(y_{2}^{2}-y_{1}^{2})y_{1}^{2}}
+\frac{1}{(r^{2}+y_{2}^{2})(y_{1}^{2}-y_{2}^{2})y_{2}^{2}}\;,\\
&&\frac{r^{2}}{(r^{2}+y_{1}^{2})(r^{2}+y_{2}^{2})}+\frac{y_{1}^{2}}{(r^{2}+y_{1}^{2})(y_{2}^{2}-y_{1}^{2})}
+\frac{y_{2}^{2}}{(r^{2}+y_{2}^{2})(y_{1}^{2}-y_{2}^{2})}=0\;,\\
&&\frac{1}{(r^{2}+y_{1}^{2})(r^{2}+y_{2}^{2})}-\frac{1}{(r^{2}+y_{1}^{2})(y_{2}^{2}-y_{1}^{2})}
-\frac{1}{(r^{2}+y_{2}^{2})(y_{1}^{2}-y_{2}^{2})}=0\;,
\end{eqnarray}
we have the following separated equations
\begin{eqnarray}
&&\frac{1}{R_{r}r^{D-6}}\partial_{r}\left(r^{D-6}\Delta_{r}\partial_{r}R_{r}\right)
+\frac{1}{\Delta_{r}}\left(r^{4}\Psi_{0}+r^{2}\Psi_{1}+\Psi\right)^{2}-\frac{a_{1}^{2}a_{2}^{2}j(j+D-7)}{r^{2}}=b_{1}r^{2}+b_{2}\;,\\
&&\left.\frac{1}{R_{y_{1}}}\left(\frac{a_{1}}{y_{1}}\right)^{D-6}\partial_{y_{1}}
\left[\left(\frac{y_{1}}{a_{1}}\right)^{D-6}(1-g^{2}y_{1}^{2})(a_{1}^{2}-y_{1}^{2})(a_{2}^{2}-y_{1}^{2}) \partial_{y_{1}}R_{y_{1}}\right]
\right.\nonumber\\
&&\ \ \ \ \ \ \ \ \ \ \ \ \ \ \ \ \ \ \ \ \ \ \ \ \ \ \ \ \ \ -\frac{1}{\Delta_{y_{1}}}\left(-y_{1}^{4}\Psi_{0}+y_{1}^{2}\Psi_{1}-\Psi_{2}\right)^{2}
-\frac{a_{1}^{2}a_{2}^{2}j(j+D-7)}{y_{1}^{2}}=b_{1}y_{1}^{2}-b_{2}\;,\nonumber\\
&&\left.\frac{1}{R_{y_{2}}}\left(\frac{a_{2}}{y_{2}}\right)^{D-6}\partial_{y_{2}}
\left[\left(\frac{y_{2}}{a_{2}}\right)^{D-6}(1-g^{2}y_{2}^{2})(a_{1}^{2}-y_{2}^{2})(a_{2}^{2}-y_{2}^{2}) \partial_{y_{2}}R_{y_{2}}\right]
\right.\nonumber\\
&&\ \ \ \ \ \ \ \ \ \ \ \ \ \ \ \ \ \ \ \ \ \ \ \ \ \ \ \ \ \ -\frac{1}{\Delta_{y_{2}}}\left(-y_{2}^{4}\Psi_{0}+y_{2}^{2}\Psi_{1}-\Psi_{2}\right)^{2}
-\frac{a_{1}^{2}a_{2}^{2}j(j+D-7)}{y_{2}^{2}}=b_{1}y_{2}^{2}-b_{2}\;,\nonumber\\
\end{eqnarray}
where $b_{1}$ and $b_{2}$ are constants.

\par In these equations the constants $\Psi_{i}$ can be obtained by considering $e^{i\psi_{0}\Psi_{0}}e^{i\psi_{1}\Psi_{1}}e^{i\psi_{2}\Psi_{2}}=e^{-i\omega t}e^{im_{1}\phi_{1}}e^{im_{2}\phi_{2}}$. Using the relationship between $t$, $\phi_{1}$, $\phi_{2}$ and $\psi_{0}$, $\psi_{1}$, $\psi_{2}$ in Eqs.~(\ref{tpsi}) to (\ref{phi2psi}), we have
\begin{eqnarray}
\Psi_{0}&=&-\omega+g^{2}(m_{1}a_{1}+m_{2}a_{2})\;,\\
\Psi_{1}&=&-\omega(a_{1}^{2}+a_{2}^{2})+m_{1}a_{1}(1+g^{2}a_{2}^{2})+m_{2}a_{2}(1+g^{2}a_{1}^{2})\;,\\
\Psi_{2}&=&-\omega a_{1}^{2}a_{2}^{2}+m_{1}a_{1}a_{2}^{2}+m_{2}a_{1}^{2}a_{2}\;.
\end{eqnarray}
More explicitly, the radial equation is
\begin{eqnarray}
&&\frac{1}{r^{D-6}}\frac{d}{dr}\left(r^{D-6}\Delta_{r}\frac{dR_{r}}{dr}\right)+\left[\frac{(r^{2}+a_{1}^{2})^{2}(r^{2}+a_{2}^{2})^{2}\omega^{2}}
{\Delta_{r}}-\frac{2\omega(1+g^{2}r^{2})(r^{2}+a_{1}^{2})^{2}(r^{2}+a_{2}^{2})^{2}}{\Delta_{r}}
\left(\frac{m_{1}a_{1}}{r^{2}+a_{1}^{2}}+\frac{m_{2}a_{2}}{r^{2}+a_{2}^{2}}\right)
\right.\nonumber\\
&&\ \ \ \ \ \left.+\frac{(1+g^{2}r^{2})^{2}(r^{2}+a_{1}^{2})^{2}(r^{2}+a_{2}^{2})^{2}}{\Delta_{r}}
\left(\frac{m_{1}a_{1}}{r^{2}+a_{1}^{2}}+\frac{m_{2}a_{2}}{r^{2}+a_{2}^{2}}\right)^{2}
-\frac{a_{1}^{2}a_{2}^{2}j(j+D-7)}{r^{2}}-b_{1}r^{2}-b_{2}\right]R_{r}=0\;,
\label{eqn:radialHT}
\end{eqnarray}
and the angular equations are, for $i=1,2$,
\begin{eqnarray}
&&\left(\frac{a_{i}}{y_{i}}\right)^{D-6}\frac{d}{dy_{i}}\left[\left(\frac{y_{i}}{a_{i}}\right)^{D-6}(1-g^{2}y_{i}^{2})(a_{1}^{2}-y_{i}^{2})
(a_{2}^{2}-y_{i}^{2})\frac{dR_{y_{i}}}{dy_{i}}\right]\nonumber\\
&&\ \ +\left\{-\frac{(a_{1}^{2}-y_{i}^{2})(a_{2}^{2}-y_{i}^{2})\omega^{2}}{1-g^{2}y_{i}^{2}}
+2\omega[m_{1}a_{1}(a_{2}^{2}-y_{i}^{2})+m_{2}a_{2}(a_{1}^{2}-y_{i}^{2})]-2m_{1}a_{1}m_{2}a_{2}(1-g^{2}y_{i}^{2})\right.\nonumber\\
&&\ \ \ \ \ \ \ \left.-\frac{m_{1}^{2}a_{1}^{2}(1-g^{2}y_{i}^{2})(a_{2}^{2}-y_{i}^{2})}
{a_{1}^{2}-y_{i}^{2}}-\frac{m_{2}^{2}a_{2}^{2}(1-g^{2}y_{i}^{2})(a_{1}^{2}-y_{i}^{2})}
{a_{2}^{2}-y_{i}^{2}}-\frac{a_{1}^{2}a_{2}^{2}j(j+D-7)}{y_{i}^{2}}
-b_{1}y_{i}^{2}+b_{2}\right\}R_{y_{i}}=0\;.\nonumber\\ \label{angulareqns}
\end{eqnarray}

\begin{figure}[t]
  \centering
  \includegraphics{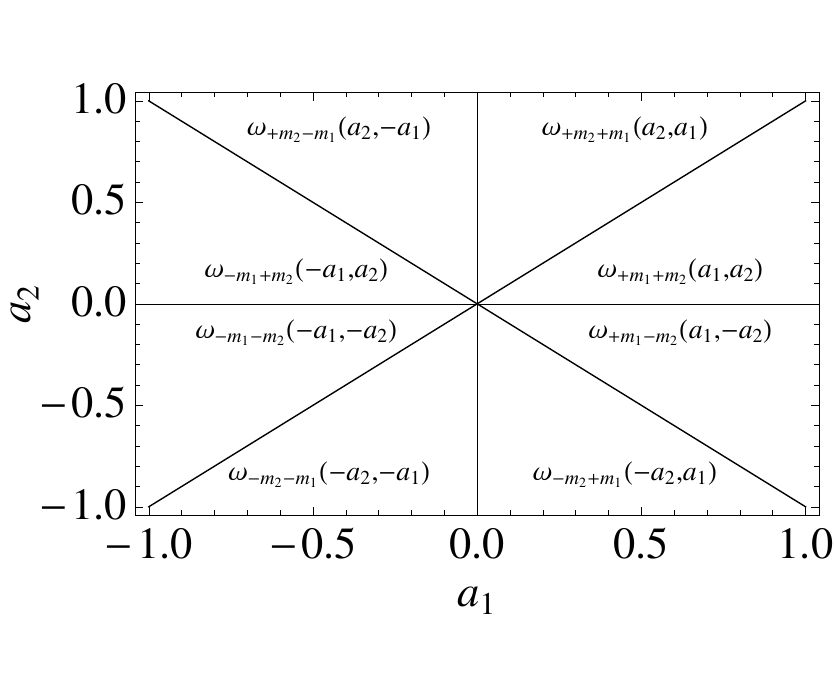}
  \caption{Due to the symmetries of the master equations we are able to patch together the QNM solutions obtained in the $a_1>a_2>0$ octant to find the solution in the whole $(a_1,a_2)$-parameter space. We are also able to find the angular eigenvalues $b_1$, $b_2$ in this way.\label{fig:symmetries_pspace}}
\end{figure}

\par These master equations possess the following  symmetries:
\begin{eqnarray}
  (m_1,a_1)\leftrightarrow(m_2,a_2)\; , \mbox{~ and ~ } (m_i, a_i)\leftrightarrow(-m_i, -a_i)\; .
\end{eqnarray}
As we shall see, because of these symmetries, we only need to calculate the QNMs and eigenvalues in the $a_1>a_2>0$ octant.  The values in the other octants can be deduced from those with their quantum numbers transformed appropriately under the above symmetries, see figure \ref{fig:symmetries_pspace}.

\par Assuming that the rotation parameters $a_{1}$ and $a_{2}$ are small, the angular eigenvalues $b_{1}$ and $b_{2}$ in the coupled angular equations above can be found perturbatively (i.e., as a power series in $\epsilon=a_{2}/a_1$ and $a_{1}$), see Appendix \ref{Ang6}. The results are:
\begin{eqnarray}
b_1&=&B_1-2\omega(m_1a_1+m_2a_2) +2g^2m_1a_2m_2a_2\;,\\
b_2&=&a_1^2B_2-2\omega(m_1a_1a_2^2+a_1^2m_2a_2)+2m_1a_1m_2a_2\;,
\end{eqnarray}
where
\begin{eqnarray}
  B_{1}&=&B_{100}+B_{102}+\mathcal{O}({\epsilon^0a_1^4,\epsilon^2})\;,
  \nn\\
  B_{2}&=& B_{200}+\left[B_{220}+B_{222}+\mathcal{O}(\epsilon^2a_1^4)\right]+\mathcal{O}(\epsilon^4)\;.
\label{smallcres}
\end{eqnarray}
The relevant terms can be obtained from equations (\ref{B200}, \ref{B100}, \ref{B102}, \ref{B220}, \ref{B222}).

%
%

\section{Numerical Method}
\label{numeric}
\par As shown in Appendix \ref{Ang6} a perturbative method can be used to determine the low order eigenvalues analytically; however, we were unable to do so in general for higher order terms, as exemplified in equation (\ref{B120}). This is unfortunate because if approximate analytic expressions in terms of $\omega$ existed for $b_1$ and $b_2$ then we could have simply substituted them into the radial equation and performed the QNM analysis without any further reference to the $y_i$ equations (at least for small values of $a_1,a_2$). The fact that the perturbative expansions of $b_1$ and $b_2$ are not algebraic expressions makes them unusable in the computation of the QNMs, because the analysis in general requires solving for the zeros of some polynomial equation in $\omega$.

\par In addition to this, we can only expect the perturbative values to hold in the small rotation regime $a_2<a_1\ll1$. Therefore, it is desirable to have an alternative method of calculating these values. In this section we will describe a method that can be used to calculate both the eigenvalues $b_1, b_2$ and the QNM, $\omega$, numerically. This will serve as a consistency check of the results obtained in Appendix \ref{Ang6} and will also allow us to go to larger values in the rotation parameters. 

\par To achieve this we will use the improved Asymptotic Iteration Method (AIM) described in \cite{Cho:2009cj}. In the current problem this method has some advantages over that of the more commonly used Continued Fraction Method (CFM) \cite{Leaver:1990zz}. In particular, the CFM requires lengthy calculations to prepare the recurrence relation coefficients that are subsequently used in the algorithm. Such manipulations are prone to error, and given the complexity of the current equations it is advantageous to use a method which bypasses this step. Furthermore, due to the existence of four Regular Singular Points (RSP) the CFM requires a further Gaussian elimination step in order to reduce the recurrence relation down to a three term recurrence relation \cite{Zhidenko:2006rs}. As we shall see the AIM works for an Ordinary Differential Equation (ODE) with four RSPs in the same way as it would for a three RSP ODE and is therefore easier to implement.\footnote{Nevertheless we found that the AIM required more iterations as the rotation parameter was increased which made it prohibitive to go beyond about $a_1\sim 1.5$. Presumably the CFM would work more efficiently in this regime, however, in the current work only the small rotation QNMs were considered.}

\par The three equations (\ref{eqn:radialHT}) and (\ref{angulareqns}) can be made to look more symmetrical by completing the square in terms of $\omega$ and defining:
\begin{eqnarray}
\label{suprad}
  \tilde{\omega}_r&=&\omega-(1+g^2 r^2) \left(\frac{m_1 a_1}{r^2+a_1^2}+\frac{m_2 a_2}{r^2+a_2^2}\right),\\
  \tilde{\omega}_{y_i}&=&\omega-(1-g^2 y_i^2) \left(\frac{m_1 a_1}{a_1^2-y_i^2}+\frac{m_2 a_2}{a_2^2-y_i^2}\right),
\end{eqnarray}
then,
\begin{eqnarray}
  0&=&\frac{1}{r^{D-6}}\frac{d}{dr}\left(r^{D-6}\Delta_{r} \frac{d R_r}{dr}\right)+\left( \frac{(r^2+a_1^2)^2(r^2+a_2^2)^2}{\Delta_r}\tilde{\omega}_r^2-\frac{a_1^2 a_2^2 j(j+D-7)}{r^2}-b_1 r^2 -b_2\right) R_r\;, \label{eqn:Rr}\\
  0&=&\left(\frac{a_i}{y_i}\right)^{ D-6} \frac{d}{dy_i}\left[\left(\frac{y_i}{a_i}\right)^{D-6}\Delta_{y_i} \frac{d R_{\theta_i}}{dy_i}\right]-\left\{ \frac{(a_1^2-y_i^2)^2(a_2^2-y_i^2)^2}{\Delta_{y_i}}\tilde{\omega}_{y_i}^2+\frac{a_1^2 a_2^2 j(j+D-7)}{y_i^2}+b_1 y_i^2 -b_2\right\} R_{\theta_i},\label{eqn:Ryi}
\end{eqnarray}
where $i=1,2$. 

\par In the next section  we shall solve the angular equations (\ref{eqn:Ryi}) showing how the AIM can be used to numerically find the $b_1$ and $b_2$ angular eigenvalues and then we shall study the radial equation (\ref{eqn:Rr}) and use the AIM to calculate the QNM, $\omega$. Before moving on, we shall briefly discuss the relation of $\tilde \omega_r$ with super-radiance and the horizon structure.

\subsubsection*{Super-radiance and the WKB form of the potential}

\par We can understand the form of $\tilde \omega_r$ when writing the radial equation in the WKB form by transforming as:
\beq
R_r(r) = r^{-D/2+3} (r^2+a_1^2)^{-1/2}(r^2+a_2^2)^{-1/2} P_r(r) \; , 
\eeq
where we defined the tortoise coordinate by
\beq
{dr_\star\over dr}=\frac{(r^2+a_1^2)(r^2+a_2^2)}{\Delta_r}\;.
\eeq
The WKB wave equation is:
\beq
\frac{d^2 P_r}{dr_\star^2}
+\left[
\tilde\omega_r^2-\frac{\Delta_r}{(r^2+a_1^2)^2(r^2+a_2^2)^2}U(r)\right]P_r=0 \, ,
\eeq
where $\tilde \omega_r$ is given in equation (\ref{suprad}) and 
\bea
U(r) &=& \Big( \frac{a_1^2 a_2^2 j (j + D - 7)}{r^2} + b_1 r^2 + b_2 \Big)- \frac{\Delta_r}{r} \Big( \frac{r^2}{(r^2 + a_1^2)^2} + \frac{r^2}{(r^2 + a_2^2)^2} - \frac{1}{2} \Big[ \frac{1}{r^2 + a_1^2} + \frac{1}{r^2 + a_2^2} \Big] \Big) \nn\\
&-& ( 1 + g^2 r^2)(r^2 + a_1^2)(r^2 + a_2^2) \Big( 3 - D/2 - \frac{r}{2}\Big[ \frac{1}{r^2 + a_1^2} + \frac{1}{r^2 + a_2^2} \Big] \Big)\Big( \frac{D - 7}{r^2} + \frac{2 g^2 }{1 + g^2 r^2} + \frac{2}{r^2 + a_1^2} + \frac{2}{r^2 + a_2^2} \Big) \nn\\
&-& \Delta_r \Big( \frac{3 - D/2}{r} - \frac{1}{2} \Big[ \frac{1}{r^2 + a_1^2} + \frac{1}{r^2 + a_2^2} \Big] \Big)^2 ~.
\eea
Given the standard solution of the WKB wave function at infinity and near the horizon (mapped to minus infinity in tortoise coordinates), this identifies $\tilde \omega_r(r_h)<0$ as super-radiant \cite{DeWitt:1975ys} for certain values of $a_1,a_2$ and $m_1,m_2$. In the super-radiant case the transmission probability ($|A|^2$) becomes negative and for Kerr-AdS we would expect this to lead to super-radiant instabilities,  see \cite{Kodama:2009rq} for an example in the simply rotating case. 

\par It may be worth mentioning that the WKB form of the radial potential could be used to find the QNMs via the WKB method of Iyer and Will \cite{iyer:1986np}, once the angular eigenvalues are known \cite{Seidel:1989bp}. It could also be used to find Hawking emissions via the WKB method \cite{Cornell:2005ux}.

\begin{figure}[t]
  \centering
  \includegraphics{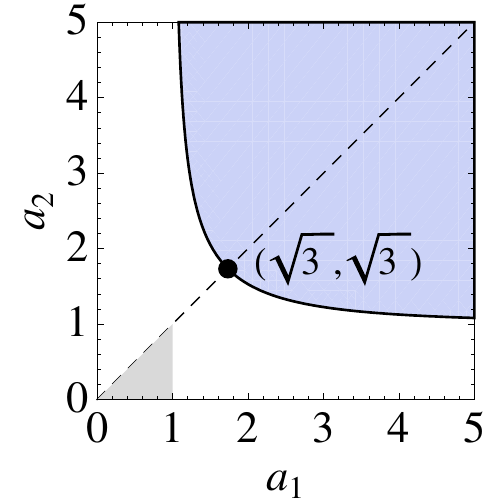}
  \caption{(Color Online) Plot of the $D=6$ parameter space. The solid curve corresponds to solutions (in units of $r_h=1$) where the two horizons overlap (degenerate). Below the curve the outer horizon is fixed to unity. As the angular momenta are increased in the direction of the curve the two horizons cross (i.e., on the degenerate curve) and then pass into the blue shaded region corresponding to solutions with the inside horizon fixed to $r_h=1$. Thus every solution below the curve (with normalisation such that the outer horizon is fixed at unity) has a corresponding equivalent solution above the curve with a different normalisation of the outside horizon. Due to symmetries we only need to study the $a_2<a_1$ region shown by the dashed line. Furthermore, in this work, we will only be investigating the small rotation $a_1\leq 1$ region shown shaded in gray. 
 \label{fig:D6pspace}}
\end{figure}

\subsubsection*{Horizon structure for $g=0$}

\par It is worth noting that even in flat space ($g=0$) the horizon structure is slightly different for $D=6$ than it is for the larger dimensions. The reason for this is related to the number of angular momentum parameters that have been set to zero. In $D=6$ there are only two possible parameters, yet in $D=7,8$ there are three and in $D=9,10$ there are four parameters. In fixing the number of rotations to only two, as we have done in the current work, $D=6$ is the only case in which the full set are present.  Furthermore, only $D=6$ and $D=7$ will have naked singularity solutions, see \cite{Doukas:2010be} for details.  

\par When obtaining our numerical results it will be necessary to fix a mass scale. Instead of setting $M=1$ it is conventional (and convenient) to set the horizon radius $r_h=1$, see for example \cite{Cardoso:2004cj}. However, setting $r_h=1$ automatically imposes the condition that a horizon exists. In this case, the solution will either have two horizons or a degenerate horizon. Using the results in \cite{Doukas:2010be}, the degenerate horizon solution can be found in the even case from the $P_2$ polynomial. In units with $r_h=1$ the degenerate solution occurs when:
\begin{equation}
  a_2^2=\frac{3+a_1^2}{a_1^2-1}\;.
\end{equation}

\par A plot of this situation is shown in figure \ref{fig:D6pspace}. As can be seen from this plot the degenerate horizon curve (solid line) divides the positive quadrant $(a_1,a_2)$-parameter space into two regions. One might wonder if the shaded blue region represents solutions with a naked singularity. However, this cannot be the case since the horizon was fixed to unity and therefore only those solutions {\it{with}} a horizon are being considered. The existence of the two regions is actually related to the fact that in general there are two horizons, an inner horizon and an outer horizon. It is the relative position of these horizons that is responsible for separating the parameter space into two. Even though we have set $r_h=1$ we have not specified which of the two horizons should take this value! Solutions under the degenerate curve in figure \ref{fig:D6pspace} correspond to those with the outer horizon fixed at unity while those above the curve correspond to solutions with the inner horizon fixed at unity. Since the two horizons are indistinguishable these two cases are really identical i.e., for every point above the line there is an equivalent solution below the line except that the normalisation of the mass is different in the two cases. Therefore, it is only necessary to study those solutions below the line in order to understand the entire stability problem.

\par In a similar way to the above reasoning it is possible to show that there are no degenerate solutions (for only two non-zero spins) in higher than six even dimensions. Furthermore, using the $P_1$ polynomial also defined in \cite{Doukas:2010be} one can again show that there are no degenerate solutions in odd dimensions\footnote{This occurs even though in $D=7$ there is a constraint on the angular momenta. In this case, the position of the degenerate horizon occurs at $r_h=0$ (this horizon has zero area and therefore should more properly be thought of as a naked singularity) and is therefore excluded by the assumption $r_h=1$ since zero can not be scaled to 1 by an appropriate choice of units.}. Thus, $D=6$ is in this sense a special case. Nevertheless, we will only be investigating the small rotation region (gray triangular region shown in figure \ref{fig:D6pspace}), and in this regime the numerical method can be implemented identically in all dimensions.

\subsection{Higher dimensional spheroidal harmonics with two rotation parameters} 
\label{subsec:AIMrotation}

\par The two equations (\ref{eqn:Ryi}) are in fact the two-rotation generalisation of the higher dimensional spheroidal harmonics (HSHs) studied in \cite{Berti:2005gp}. In this case, the existence of two rotation parameters leads to a system of two coupled second order 
ODEs\footnote{For the moment we are considering $\omega$ to be an independent parameter.}. We note that, in general, one would expect that the generalisations of the HSHs to $\lfloor \frac{D-1}{2}\rfloor$ rotation parameters would lead to even larger systems of equations. While these systems would also be useful generally in studies of MP black holes, here we will only focus on the two rotation case.

\par  It can be seen that these equations have regular singular points at $y_i^2=a_1^2, a_2^2, \tfrac{1}{g^2}$ and $0$. We assume that the cosmological constant is small and in particular that $a_1^2\ll|\tfrac{1}{g^2}|$. Recall that $y_1$ and $y_2$ are defined on the domains shown in constraint (\ref{yconstraints}). We would therefore expect the solutions to be well-behaved except possibly at the boundaries of these domains where singularities are present. In order to determine the regular solutions we need to define an appropriate norm on the space of solutions. First we change to the variable $y_i^2=\xi_i$. The angular equations can then be written in the Sturm-Liouville form (assuming momentarily that $\omega$ and $b_2$ are real):
\begin{equation}
  \lambda w(\xi_i) R_{\theta_i}(\xi_i)=-\frac{d}{ d\xi_i}\left(p(\xi_i) \frac{d}{d\xi_i}R_{\theta_i}(\xi_i)\right)+q(\xi_i)R_{\theta_i}(\xi_i)
\end{equation}
with the weight function $w_1(\xi_i)=\tfrac{1}{4} \xi_i^{(D-5)/2}$, the eigenvalue $\lambda=-b_1$, and 
\begin{eqnarray}
  p(\xi_i)&=&\xi_i^{(D-5)/2}\Delta_{\xi_i},\\
  q(\xi_i)&=&\frac{1}{4} \xi_i^{(D-7)/2}\left(\frac{(a_1^2-\xi_i)^2(a_2^2-\xi_i)^2}{\Delta_{\xi_i}}\tilde{\omega}_{\xi_i}^2+\frac{a_1^2 a_2^2 j(j+D-7)}{\xi_i} -b_2\right)\; ,
\end{eqnarray}
where $\Delta_{\xi_i}$ and $\tilde{\omega}_{\xi_i}$ are defined in the obvious way under the change of coordinates. Since $w(\xi)>0$ we can define the two norm's:
\begin{eqnarray}
  N_1(R_{\theta_1})&\propto&\int^{a_1^2}_{a_2^2}\xi_1^{(D-5)/2} |R_{\theta_1}|^2 d\xi_1\;,\\
  N_2(R_{\theta_2})&\propto& \int^{a_2^2}_{0}\xi_2^{(D-5)/2} |R_{\theta_2}|^2 d\xi_2\;.
\end{eqnarray}
The rationale for this choice can be explained as follows. We can rewrite the Sturm-Liouville equation as an eigenvalue equation $L R=\lambda R$ where 
\begin{equation}
  L=\frac{1}{w(\xi)} \left(-\frac{d}{d\xi}\left[p(\xi) \frac{d}{d\xi}\right]+q(\xi)\right).
\end{equation}
In analogy to the criterion discussed in \cite{Kodama:2009rq} we note that for real $\omega$ and real $b_2$, $\lambda$ (or $b_1$) must be real. Therefore the inner product must be chosen so that $L$ is self-adjoint when $\omega$ and $b_2$ are real. From the Sturm-Liouville form it is easy to show\footnote{With appropriate boundary conditions.} that if the inner product is defined as:
\begin{equation}
  \langle f,g \rangle = \int f^*(\xi)g(\xi) w(\xi) d\xi\;,
\end{equation}
then $L$ is self-adjoint i.e., $\langle f , L g \rangle= \langle L f, g\rangle$. This inner product naturally induces the norms chosen above. 

\par However, one readily sees that the choice of norm is not unique. We could for example repeat the argument made above using $\lambda=b_2$ and in this case the weight function is found to be $w_2 (\xi_i)=\frac{1}{4} \xi_i^{(D-7)/2}$. The main point, however, is that even though the norms will give a different number (when acting on a given solution), they will agree on which solutions are regular (finite norm) \footnote{In Appendix \ref{Ang6} this ambiguity is somewhat more relevant. We find that in order to be able to simplify the expressions using the Jacobi orthonormality relations, one must choose the $w_1$ to normalize the $R_{\theta_1}$ solutions and $w_2$ to normalize the $R_{\theta_2}$ solutions respectively.}.

Under either choice of weight the regular solutions are found to be:
\begin{eqnarray}\label{eqn:modeR1}
  R_1&\sim& (\xi_1-a_2^2)^{\frac{|m_2|}{2}}(a_1^2-\xi_1)^{\frac{|m_1|}{2}}\Psi_1;\quad \xi_1\in (a_2^2,a_1^2),\\
  R_2&\sim& \xi_2^{j/2}(a_2^2-\xi_2)^{\frac{|m_2|}{2}}\Psi_2;\quad \xi_2\in (0,a_2^2).\label{eqn:modeR2}
\end{eqnarray}
Now for a given value of $\omega$ we can determine $b_1$ and $b_2$ simply by performing the improved AIM \cite{Cho:2009cj} on both of the angular equations separately. This will result in two equations in the two unknowns $b_1,b_2$ which we can then solve using a numerical routine such as the built-in Mathematica functions NSolve or FindRoot. More specifically we rewrite equations (\ref{eqn:Ryi}) using (\ref{eqn:modeR1}) and (\ref{eqn:modeR2}) and transform them into the AIM form:
\begin{eqnarray}
  \frac{d^2\Psi_1}{d \xi_1 ^2}&=&\lambda_{01} \frac{d\Psi_1}{d \xi_1}
+s_{01} \Psi_{1}\;,\\
  \frac{d^2\Psi_2}{d \xi_2 ^2}&=&\lambda_{02} \frac{d\Psi_2}{d \xi_2}
+s_{02} \Psi_{2}\;.
\end{eqnarray} 
The AIM requires that a special point be taken about which the $\lambda_{0i}$ and $s_{0i}$ coefficients are expanded. As was shown in \cite{Cho:2009wf} different choices of this point can worsen or improve the speed of the convergence. In the absence of a clear selection criterion we simply choose this point conveniently in the middle of the domains:
\begin{eqnarray}
  \xi_{01}=\frac{a_1^2+a_2^2}{2},\quad
  \xi_{02}=\frac{a_2^2}{2}\;.
\end{eqnarray}


\subsubsection*{Eigenvalue results and comparison with double perturbation theory}

\begin{figure}[t]
  \centering
  \includegraphics[scale=.8]{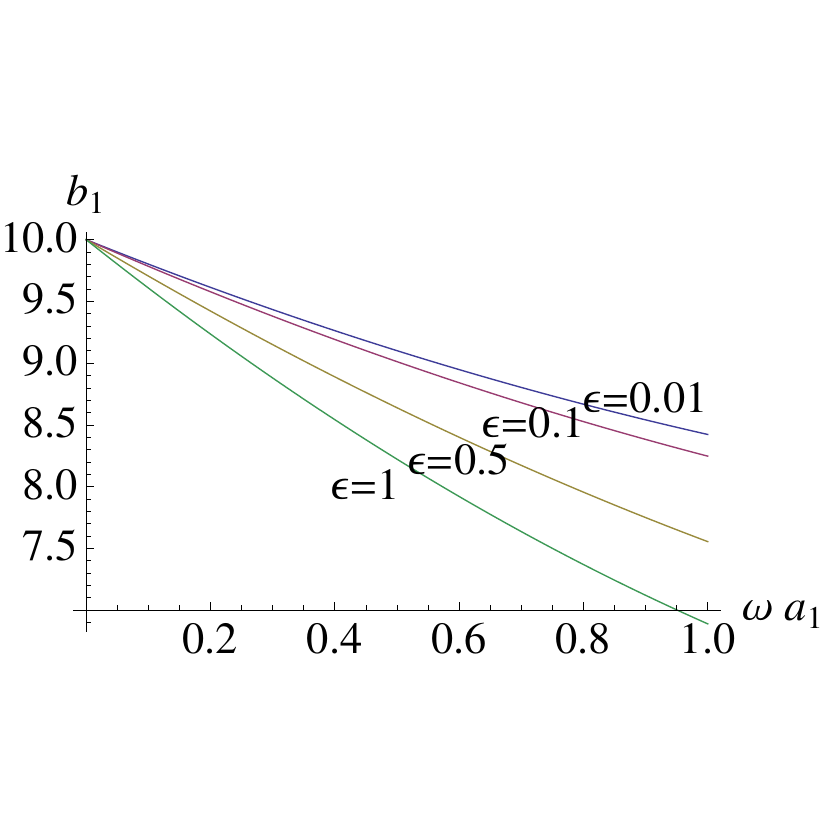}
  \includegraphics[scale=.8]{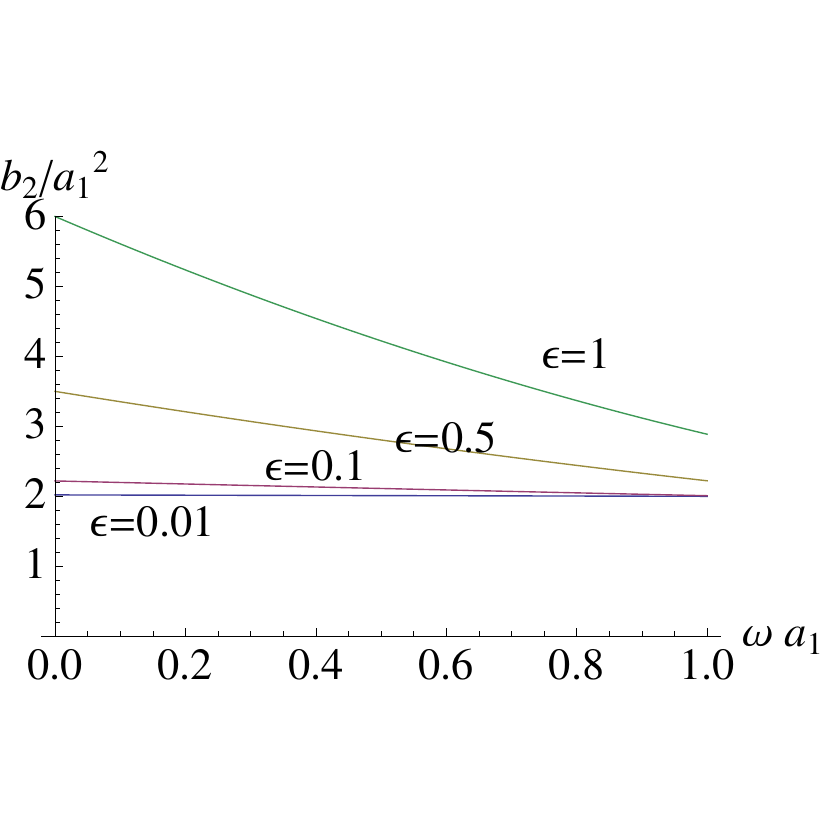}
  \caption{(Color Online) $D=6$, $g=0$, $(j,m_1,m_2,n_1,n_2)=(0,1,1,0,0)$. A plot of the eigenvalues for various choices of $\epsilon\equiv a_2/a_1$. Note that the dependence on $a_1$ has been scaled into the other quantities. \label{fig:evalepsilon}}
\end{figure}
\begin{figure}[t]
  \centering
  \includegraphics[scale=.8]{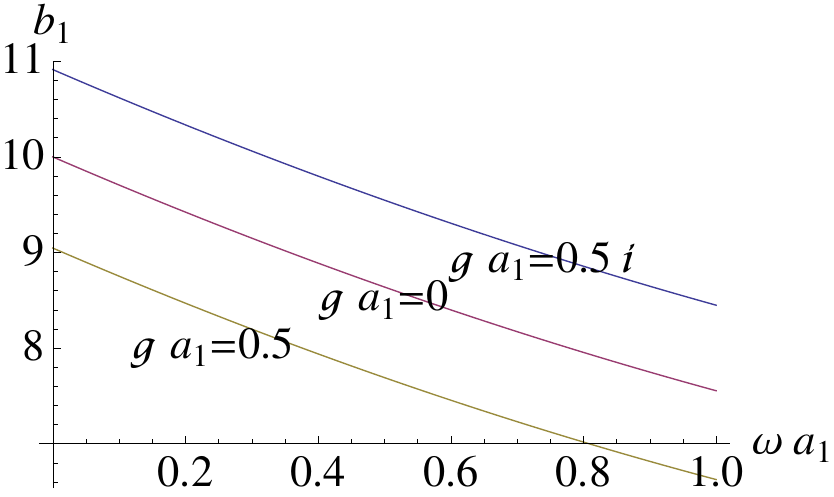}
  \includegraphics[scale=.8]{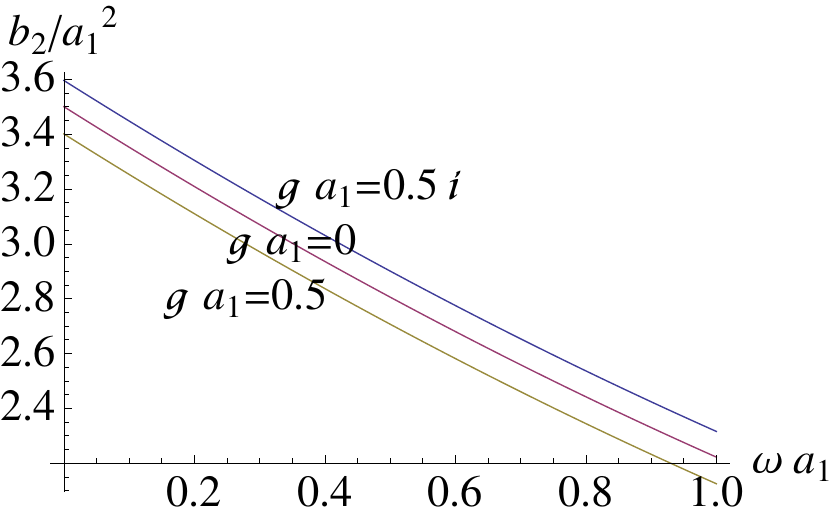}
  \caption{(Color Online) $D=6$, $\epsilon\equiv a_2/a_1=1/2$,  $(j,m_1,m_2,n_1,n_2)=(0,1,1,0,0)$. A plot of the eigenvalues for $g a_1 = 0.5 i, 0, 0.5$, corresponding to deSitter, flat, and anti-deSitter spacetimes respectively. Note that the dependence on $a_1$ has been scaled into the other quantities. \label{fig:evalg}}
\end{figure}

\par The numerical $b_1$ and $b_2$ eigenvalues for various parameters were computed and shown to be in good agreement with the perturbative values, Appendix \ref{Ang6}, for small $\epsilon$ and $\omega$. This serves as a consistency check between these two methods. Some results are plotted in figures \ref{fig:evalepsilon}, \ref{fig:evalg}. However, some issues arose and we found that both methods had their limitations, which we now briefly outline.

\begin{figure}[h]
  \centering
  \includegraphics[scale=0.8]{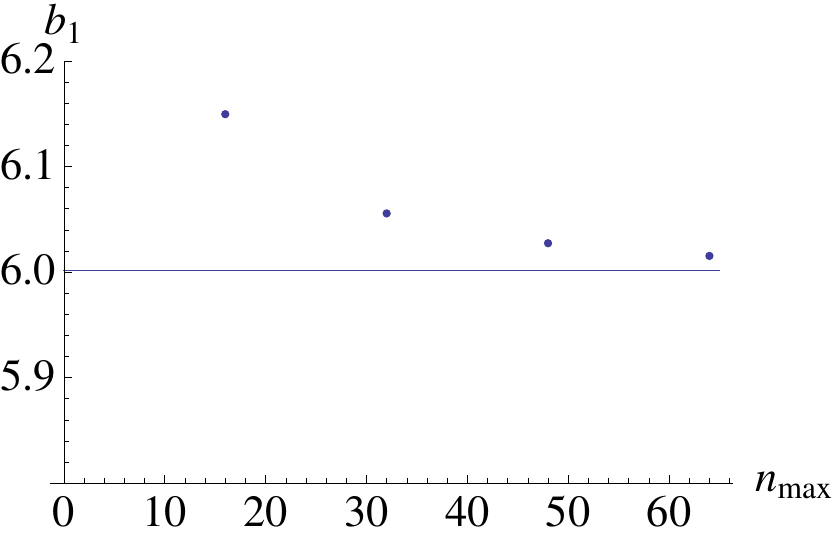}
  \caption{(Color Online) $D=6$, $a_1=3/2$, $a_2=149/100$, $(j,m_1,m_2,n_1,n_2)=(0,1,1,0,0)$ and $g=0$. The dots are the numerical $b_1$ eigenvalues for an increasing number of AIM iterations, while the straight line is the value obtained from the perturbative method to $\mathcal{O}(\epsilon^6, a_1^6)$. \label{fig:epsilon1} 
 }
\end{figure}

We found that for $\epsilon\ll1$ there was no appreciable difference between the numerical eigenvalues found after $16$ or $32$ iterations. In other words the convergence was quite fast. As $\epsilon\rightarrow 1$ however we found that the convergence was much slower. For example, at $a_1=3/2$, $a_2=149/100$ (i.e., $\epsilon=0.99\dot{3}$) we needed about $80$ iterations to get to the same level of accuracy that we required for smaller epsilon. See figure \ref{fig:epsilon1}. In this case (for small values of $\omega$) the perturbative method outperformed the AIM.

\begin{figure}[h]
  \centering
  \includegraphics[scale=0.8]{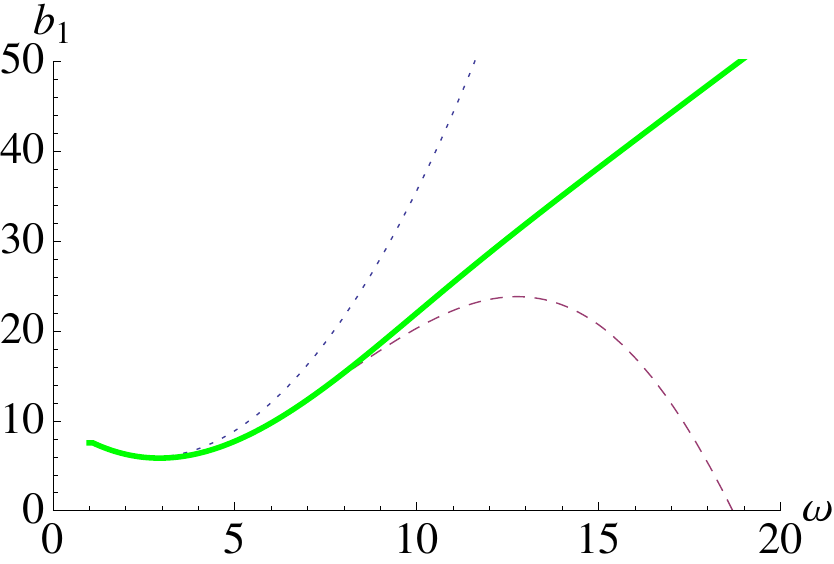}
  \caption{(Color Online) $D=6$, $a_1=1$, $a_2=1/2$, $(j,m_1,m_2,n_1,n_2)=(0,1,1,0,0)$ and $g=0$. The dotted blue line is the perturbation to order $\mathcal{O}(\epsilon^2, a_1^2)$, while the dashed red line is the perturbation to order $\mathcal{O}(\epsilon^6, a_1^6)$. The solid green line is the converged numerical $b_1$ eigenvalue. 
 }
  \label{fig:largeomega} 
\end{figure}

However, we also found that the perturbative eigenvalues were very poor as $\omega$ became large. As an example, we choose the point $a_1=1, a_2=1/2$. Since $\epsilon=1/2$ was relatively small we again found only $16$ AIM iterations were required to get numerical convergence. However, with $\omega>5$ the perturbative values were clearly breaking down, see figure \ref{fig:largeomega}.

\par The reason for this is that essentially the perturbation is expanded in the parameter $\omega a_1$ where $\omega$ is assumed to be order unity. However, if $\omega$ is large then the error in this expansion becomes worse.

\par The fact that the convergence of the numerical method is not sensitive to $\omega$ makes it more robust when calculating the QNMs, especially since we were only able to obtain the eigenvalues in the general case to $\mathcal{O}(\epsilon^2,a_1^2)$. Of course, what we have learned is that if we hold the number of AIM iterations at $16$ then we would expect any values we calculate to have larger errors as $\epsilon$ increases. In the next section we will calculate the QNMs completely numerically.


\subsection{Radial quasi-normal modes}

\par Having successfully determined two independent methods for calculating the eigenvalues and having showed that they agree well with each other we can now proceed to calculate the QNMs with control over their range of validity. We start with the radial master equation (\ref{eqn:Rr}). Thus far, when calculating the eigenvalues, we have been able to work in full generality, i.e., including the cosmological constant. However, due to the presence of new horizons and different boundary conditions, the flat, deSitter and anti-deSitter QNMs will need to be calculated separately. In this work we will focus only on the flat case, and from here on we set $g=0$.

\par Recall that QNMs are solutions to the radial master equation which satisfy the boundary condition that there are only waves ingoing at the black hole horizon and outgoing at asymptotic infinity. However, we found the AIM seems to work best on a compact domain. Therefore it is better to define the variable $x=1/r$, so that infinity is mapped to zero and the outer horizon stays at $x_h=1/r_h=1$. The domain of $x$, therefore, will be $[0,1]$. Thus the QNM boundary condition is translated into the statement that the waves move leftward at $x=0$ and rightward at $x=1$.  We again choose the AIM point in the middle of the domain, i.e., at $x=1/2$.

\par In terms of $x$ the radial equation (\ref{eqn:Rr}) becomes:
\begin{equation}\label{eqn:radialx}
  0=-x^{D-4}\frac{d}{dx}\left(-x^{8-D}\Delta_{x} \frac{d R}{dx}\right)+\left( \frac{(x^{-2}+a_1^2)^2(x^{-2}+a_2^2)^2}{\Delta_x}\tilde{\omega}_x^2-a_1^2 a_2^2 j(j+D-7)x^2-\frac{b_1}{x^2} -b_2\right) R\;, 
\end{equation}
where $\Delta_x(x)\equiv\Delta_r(r=1/x)$ and $\omega_x(x)\equiv\omega_r(r=1/x)$.

\par After performing some asymptotic analysis, keeping in mind the definition (\ref{eqn:SeparationAnsatz}), we find that for the solutions to satisfy the QNM boundary conditions we must have:
\begin{equation}\label{eqn:y}
  R \sim (1-x)^{i \tilde{\omega}_h\alpha_h} x^{(D-2)/2}e^{i \omega_x/x} y(x)\;, 
\end{equation}
where 
\begin{eqnarray}
  \tilde{\omega}_h&\equiv&\omega_x(x=1)\;,\\
  \alpha_h&\equiv&\frac{(1+a_1^2)(1+a_2^2)}{\Delta_x'(x=1)}\;.
\end{eqnarray}
We then substitute this ansatz into equation (\ref{eqn:radialx}) and rewrite into the AIM form:
\begin{equation}
y''=\lambda_0 y' +s_0 y\;.
\end{equation}
This final step was performed in Mathematica and then the resulting expressions for $\lambda_0$ and $s_0$ were fed into the AIM routine.

\begin{figure}[t]
  \centering
  \includegraphics{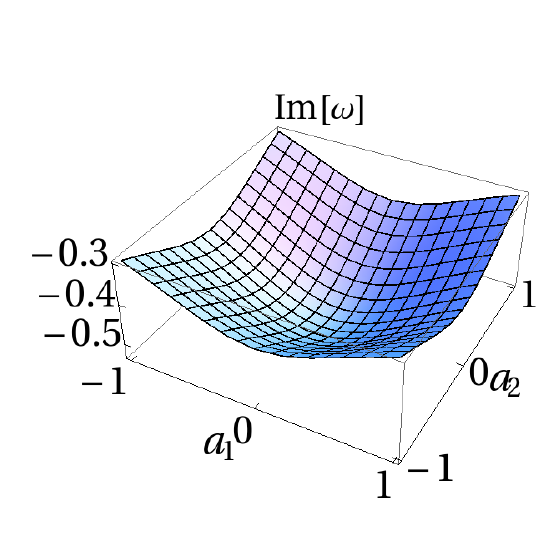}
  \includegraphics{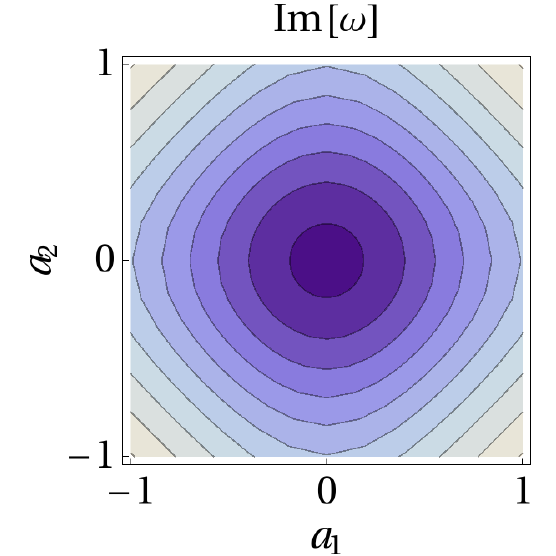}
  \includegraphics{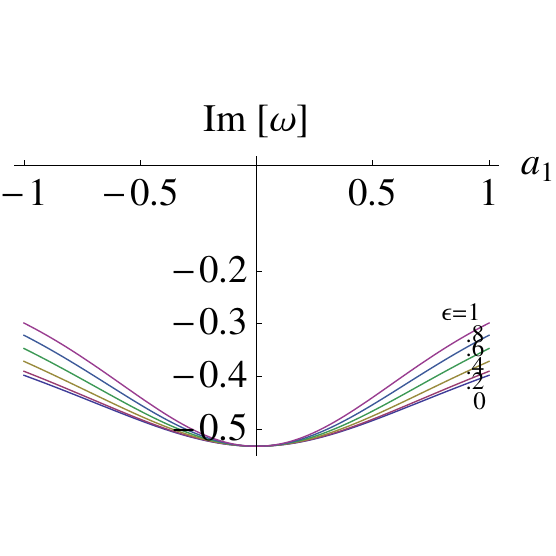}
  \includegraphics{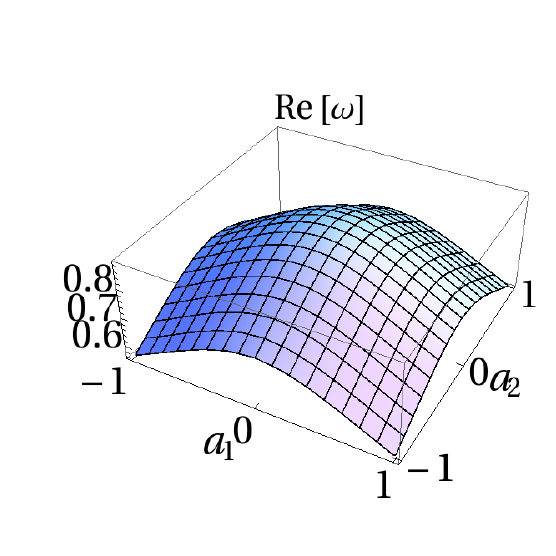}
  \includegraphics{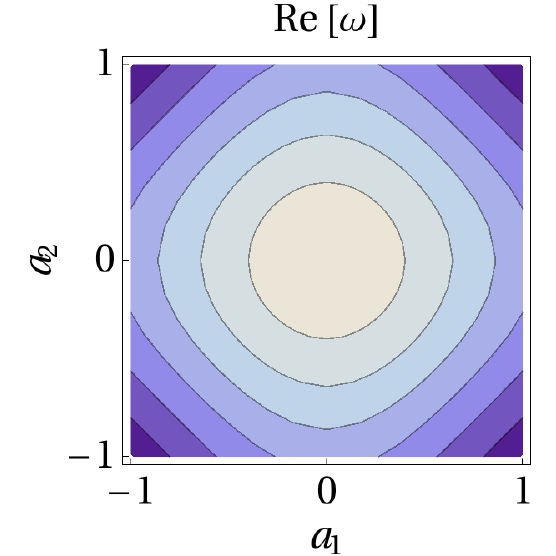}
  \includegraphics{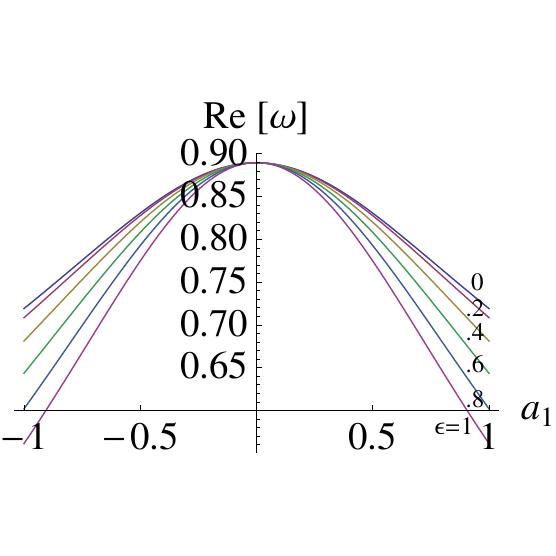}
\caption{(Color Online) D=6. Plots of the fundamental $(j,m_1,m_2,n_1,n_2)=(0,0,0,0,0)$ QNM. On top are plots of the imaginary part and below are plots of the Real part. Left: a surface plot over the $(a_1,a_2)$-parameter space. Middle: a contour plot. Right: a plot of the QNM along the straight lines passing through the origin with gradient $\epsilon$ shown in the graph. 
\label{fig:D6000plots}}
\end{figure}

\begin{figure}[t]
  \centering
  \includegraphics{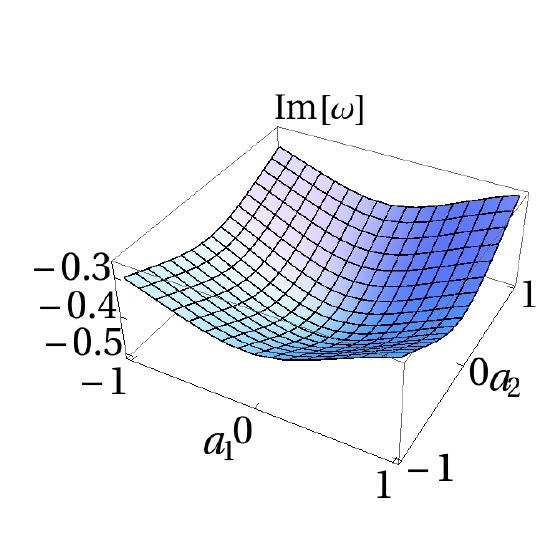}
  \includegraphics{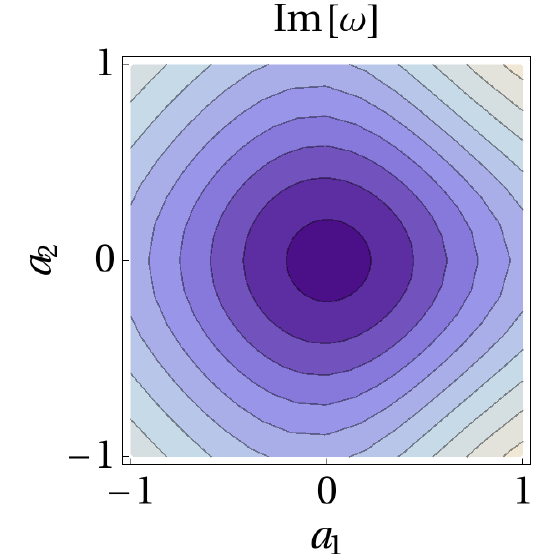}
  \includegraphics{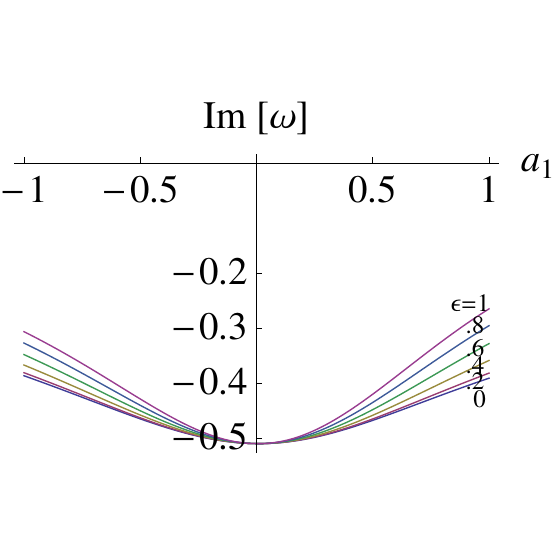}
  \includegraphics{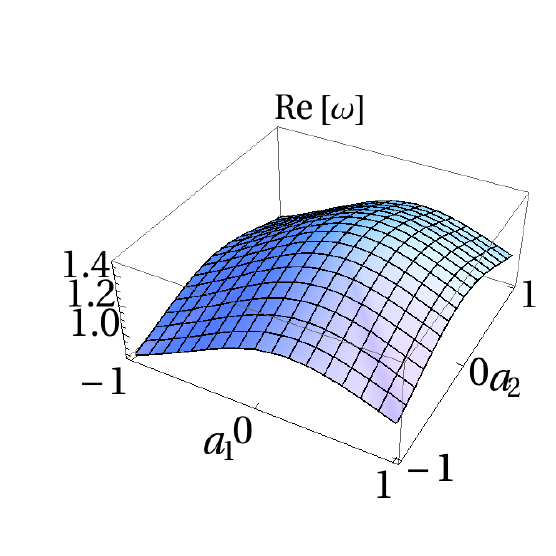}
  \includegraphics{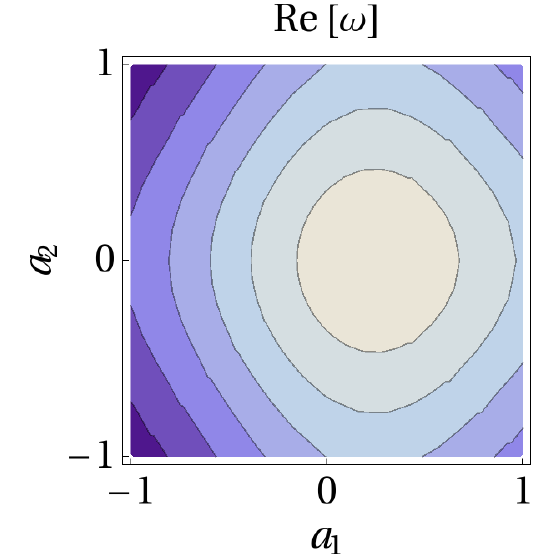}
  \includegraphics{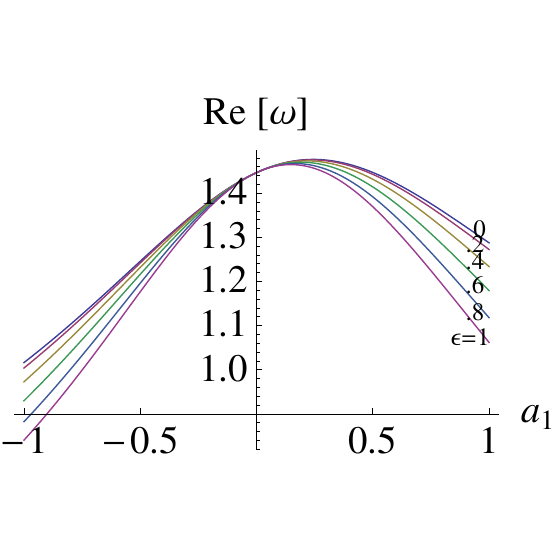}
  \caption{(Color Online) D=6. Plots of the fundamental $(j,m_1,m_2,n_1,n_2)=(0,1,0,0,0)$ QNM. On top are plots of the imaginary part and below are plots of the Real part. Left: a surface plot over the $(a_1,a_2)$-parameter space. Middle: a contour plot. Right: a plot of the QNM along the straight lines passing through the origin with gradient $\epsilon$ shown in the graph. 
  \label{fig:D6010plots}}
\end{figure}

\par The method we use to find the QNMs proceeds in a fashion similar to that used in \cite{Berti:2005gp, PhysRevD.73.064030}. However, as already mentioned we use the AIM instead of the CFM.
\par First we set the number of AIM iterations in both the eigenvalue and QNM calculations to sixteen\footnote{Ideally this should be made as large as possible, however, we found that there was little difference in the computed QNMs when using 16 or 32 iterations for $a_1\leq1$. This choice also gave good agreement in the small rotation regime with the perturbed eigenvalues calculated in appendix \ref{Ang6}. However, to go into the large rotation limit we found that much larger iterations were required to achieve convergence and this significantly slowed down the code. Thus the AIM method we have described here does not seem robust enough to explore the large rotation limits.}. We start with the Schwarzschild values $(b_1,b_2,\omega)$, i.e., at the point $(a_1,a_2)\sim 0$\footnote{Note we couldn't take the point (0,0) exactly as this would leave the $y_1$ and $y_2$ domains empty. To be precise we chose the point $(a_1,a_2)=(1/50,1/100)$.}  and then increment $a_1$ and $a_2$ by some small value\footnote{In our results we incremented $a_1$ by $\frac{1}{50}$ at each step. The increment in $a_2$ was then set by the gradient of the straight line taken in the $(a_1,a_2)$- parameter space}. We take the initial eigenvalues $(b_1,b_2)$, insert them into the radial equation (\ref{eqn:y}) then use the AIM to find the new QNM that is closest to $\omega$ using the Mathematica routine FindRoot. 

\par We then take this new value of omega, $\omega'$, insert it into the two angular equations (at the same value of $a_1$ and $a_2$) then solve using the AIM and searching closest to the previous $b_1$ and $b_2$ values. Thereby obtaining the new eigenvalues $b_1'$, $b_2'$. We then repeat this process with the new $(\omega',b_1',b_2')$ as the starting point until the results converge and we have achieved four decimal places of accuracy\footnote{We found that no more than 15 repetitions were required to achieve convergence.}. When this occurs we increment $a_1$ and $a_2$ again and repeat the process. In this way, we are able to find the QNMs and eigenvalues along lines passing approximately through the origin (i.e., starting from the near Scwharzschild values) in the $(a_1,a_2)$ parameter space. We choose 6 straight lines with gradients of $(\epsilon=0,0.2,0.4,0.6,0.8,1)$\footnote{Note that if these lines went exactly through the origin, $\epsilon=0$ would mean $a_2=0$ and $\epsilon=1$ would mean $a_2=a_1$ both of these situations would be pathological to the numerical method since the $y_i$ domains would disappear. With the starting value at $(1/50,1/100)$, we come very close to the single rotation and $a_1=a_2$ cases for $\epsilon=0$ and $\epsilon=1$ respectively while remaining in a valid domain of the numerical procedure.}  and then use an interpolating function to interpolate the values in between these points. This then covers the $a_1>a_2>0$ octant. 

\par Some preliminary results are shown for $D=6$, in figures \ref{fig:D6000plots} and \ref{fig:D6010plots}. We see that as we increase the gradient the imaginary part of the curves appear to be bounded between the $a_2=0$ to $a_1=a_2$ curves. Indeed, if this behaviour is a general phenomenon, then the most important regime for locating instabilities (i.e., when the imaginary part crosses the $Im (\omega)=0$ axis) would appear to be the $a_1=a_2$ limit. Some general features of these plots are worth mentioning. For the case of vanishing angular modes $m_{1,2}=0$, the solution is symmetric under horizontal and vertical reflections in the $a_1$ and $a_2$ axes in figure \ref{fig:D6000plots}. Furthermore, with $m_1$ non-zero this reflection symmetry is broken in the real part of the QNM curves shown on the bottom right of figure \ref{fig:D6010plots}, where they are skewed to the left. We have also confirmed this the case for $m_2\neq 0$.

\par
\begin{figure}[t]
  \centering
  \includegraphics{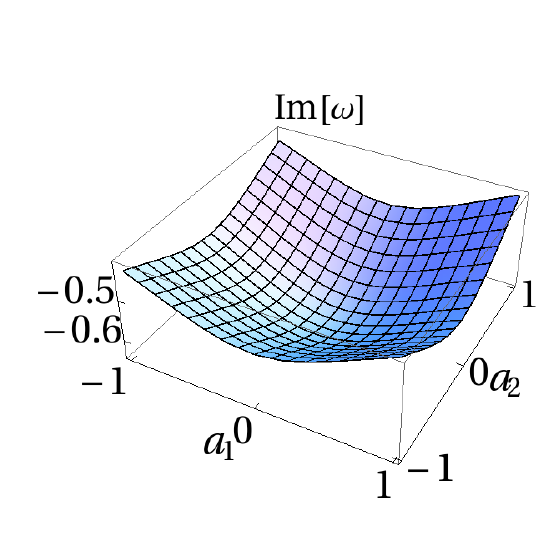}
  \includegraphics{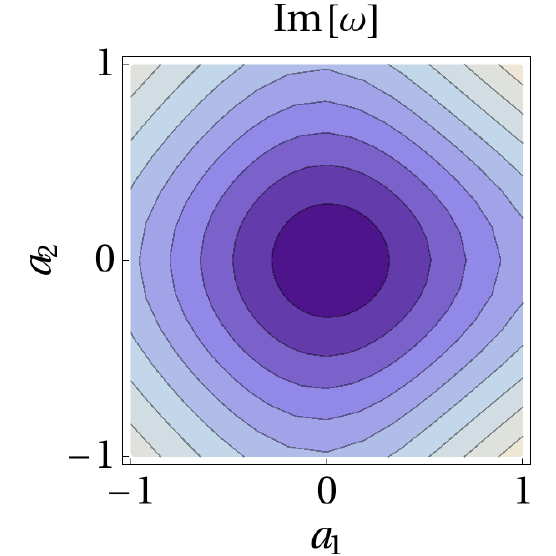}
  \includegraphics{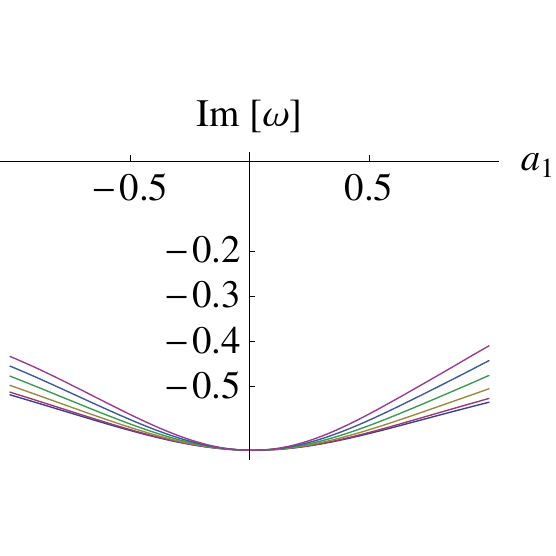}
  \includegraphics{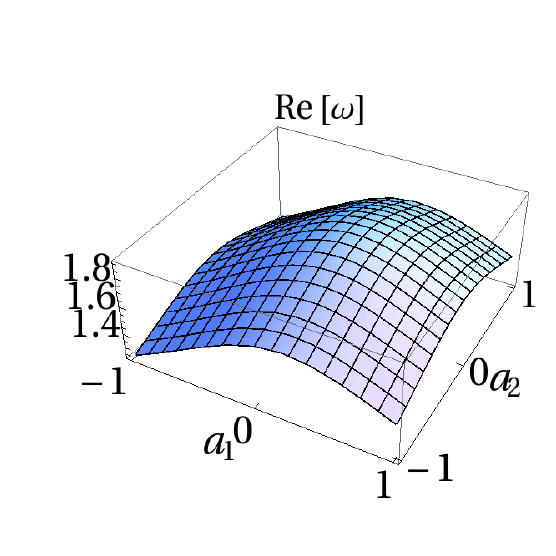}
  \includegraphics{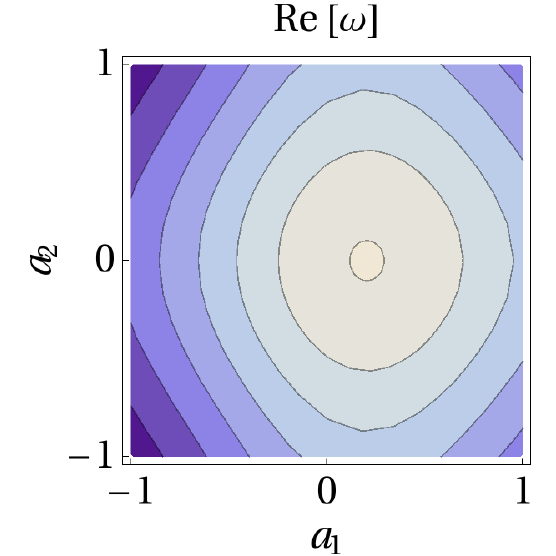}
  \includegraphics{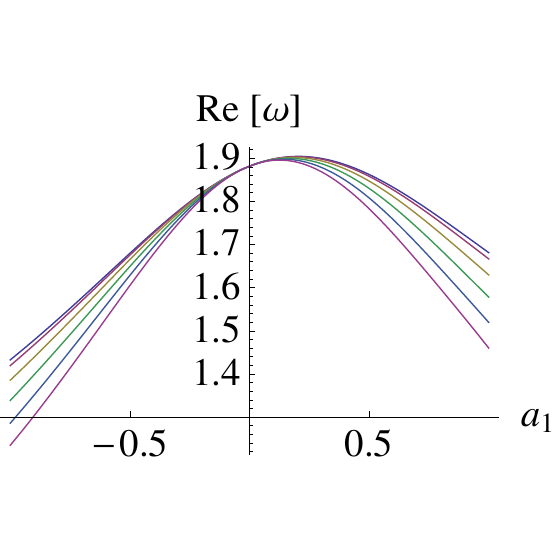}
  \caption{(Color Online) D=7. Plots of the fundamental $(j,m_1,m_2,n_1,n_2)=(0,1,0,0,0)$ QNM. On top are plots of the imaginary part and below are plots of the Real part. Left: a surface plot over the $(a_1,a_2)$-parameter space. Middle: is a contour plot. Right: is a plot of the QNM along the straight lines passing through the origin with gradient $\epsilon$ shown in the graph. 
  \label{fig:D7010plots}
  }
\end{figure}

\begin{figure}[t]
  \centering
  \includegraphics{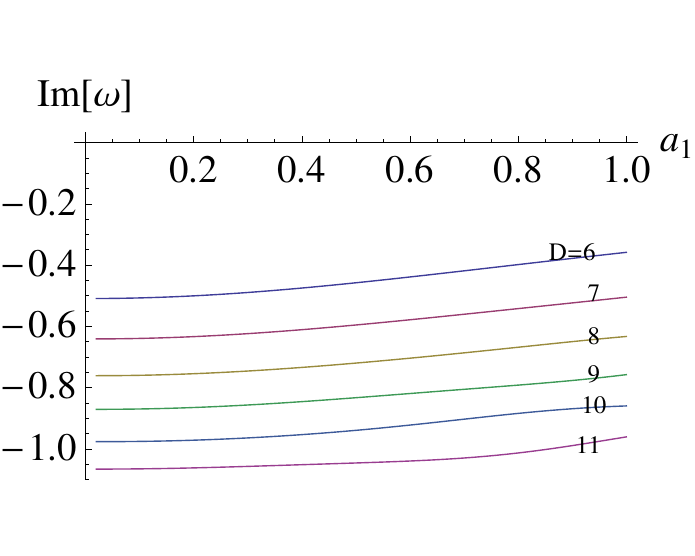}
  \hspace{1cm}
  \includegraphics{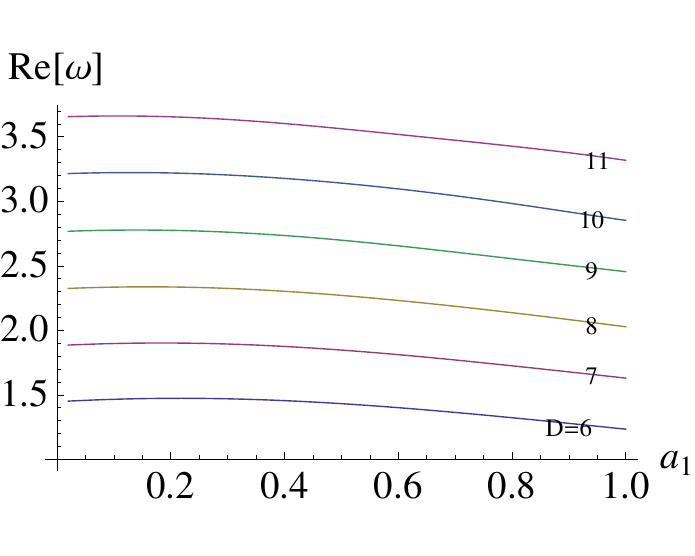}
  \caption{(Color Online) Dimensional dependence of fundamental $(j,m_1,m_2,n_1,n_2)=(0,1,0,0,0)$ QNM for $a_2=0.4 a_1$ for $0<a_1<1$. 
  \label{fig:dimensionPlots}
  }
\end{figure}

 For higher dimensions such as $D=7$, see figure \ref{fig:D7010plots}, we observe similar behaviour to the $D=6$ case where again we see the skewing of the real part for non-zero $m_1$, for example. The general dimensional dependence is shown in figure \ref{fig:dimensionPlots}, where as is typical of singly rotating cases, larger dimensions lead to greater negative $Im (\omega)$ implying larger damping. These results also seem to indicate that larger values of $D$ are more stable with increasing $a_1$.

%
%

\section{Concluding remarks}
\label{conc}

\par The intention of this work was to initiate the study of higher-dimensional Kerr-(A)dS black holes with with more than one rotation parameter for $D>5$. In the present work we have considered those solutions with all rotation parameters set to zero except for two. As a first step in this direction, we have presented the general metric for such a spacetime. We have also separated the Klein-Gordon equation, writing out the corresponding radial and angular equations explicitly. In the general case with all rotations, e.g., see \cite{Chen:2006xh,Gibbons:2004uw,Gibbons:2004js}, the separation must be performed for each $D$ separately. However, we found that in the doubly rotating case ($a_1\neq 0, a_2\neq0, a_{3}=a_4=\cdots=0$) a general $D$-dimensional expression could be obtained analogous to the one commonly used in the simply rotating case.

\par It is worth stressing that in five dimensions there is only one one spheroidal equation, while in six and higher dimensions there are two angular equations, therefore in this work we only focussed on the $D\geq6$ case (the five dimensional case will be considered elsewhere \cite{Cho:inPROG, Aliev:2008yk}). We evaluated the QNM frequencies of the low-lying modes using a numerical AIM approach for both the angular and radial equations. In Appendix \ref{Ang6}, to get some quantitative understanding of the angular equations, we also developed perturbative expansions in powers of the rotation parameters $\epsilon=a_{2}/a_{1}$ and $a_{1}$ for the angular eigenvalues.

\par Our preliminary results for the QNMs suggest that slowly rotating black holes with two rotations are stable although our numerical code became slow for values of $a_1,a_2\geq 1$, which unfortunately is also the region of most concern. More work in this direction, particularly for larger rotations and $g>0$ (Kerr-AdS), would also be worthy of investigation.

\par On the other hand, to discuss the stability of the ultra-spinning simply rotating black holes the angular eigenvalue in the large, imaginary, rotation limit \cite{Berti:2005gp}, is typically relevant. For our case there are two rotation parameters. With one rotation parameter small and the other large, the situation will be very much like the simply rotating case \cite{Morisawa:2004fs, Cardoso:2004cj} and no instability is expected. Therefore it would more interesting to consider the other case with both rotation parameters large. This work will be pursued subsequently.
 
\par In terms of other future work, with these separated equations we could also start to ask questions about the spectra of Hawking radiation (for five dimensions see \cite{Nomura:2005mw}) and investigate the phenomenon of super-radiance in more detail.

%
%

\acknowledgments

\par  HTC was supported in part by the National Science Council of the Republic of China 
under the Grant NSC 99-2112-M-032-003-MY3, and the National Science Centre for 
Theoretical Sciences. The work of JD was supported by the 
Japan Society for the Promotion of Science (JSPS), under fellowship no. P09749.

\appendix

%
%

\section{Double Perturbation Theory}
\label{Ang6}

\par In this appendix we develop expansions in powers of the parameters $\epsilon=a_{2}/a_1$ and $a_{1}$ for the angular separation constants $b_1$ and $b_2$ defined in the equations (\ref{angulareqns}). Since without loss of generality we have taken $a_1>a_2$, $\epsilon<1$, furthermore, we assume $a_1<1$.

\par For brevity we shall outline the main steps and refer the reader to reference \cite{Cho:inPROG} for a more detailed account of the five-dimensional case\footnote{For $D=5$ there are two rotation parameters, but there is only one angular equation. This makes the perturbative analysis easier.}. Since the latitude coordinates are restricted to $a_{2}\leq y_{1}\leq a_{1}$ and $0\leq y_{2}\leq a_{2}$, it is convenient to change variables to $x_{1}$ and $x_{2}$ with
\begin{eqnarray}
y_{1}^{2}=\frac{1}{2}\left(a_{1}^{2}+a_{2}^{2}\right)-\frac{1}{2}\left(a_{1}^{2}-a_{2}^{2}\right)x_{1}\ \ \ \ \ ;\ \ \ \ \ y_{2}^{2}=\frac{1}{2}a_{2}^{2}\left(1-x_{2}\right),
\end{eqnarray}
where $-1\leq x_{1},x_{2}\leq 1$.

\par By making the convenient choice
\begin{eqnarray}
B_{1}&\equiv&b_{1}+2\omega(m_1 a_{1}+m_2 a_{2})-2g^{2}m_1 a_{1}m_2 a_{2},\\
B_{2}&\equiv&\frac{1}{a_{1}^{2}}\left[b_{2}+2\omega(m_1a_{1}a_{2}^{2}+a_{1}^{2}m_2a_{2})-2m_1a_{1}m_2a_{2}\right]
\label{smallbs}
\end{eqnarray}
in Eq.~(\ref{angulareqns}) the perturbative expansion then only develops even powers of $\epsilon$ and $a_1$. In this case the operator equations are:
\begin{eqnarray}
{\cal O}_{1}&=&-\frac{1+x_{1}}{2}\left[(1-x_{1})+\epsilon^{2}(1+x_{1})\right]
\left\{2-g^{2}a_{1}^{2}\left[(1-x_{1})+\epsilon^{2}(1+x_{1})\right]\right\}\frac{d^{2}}{dx_{1}^{2}}\nonumber\\
&&+\left\{\left[\frac{1}{2}\left((D-5)+(D-1)x_{1}\right)
-\frac{1}{4}g^{2}a_{1}^{2}(1-x_{1})\left((D-3)+(D+1)x_{1}\right)\right]\right.\nonumber\\
&&\ \ -\frac{\epsilon^{2}}{2}\left(\frac{1+x_{1}}{1-x_{1}}\right)
\left[\left((D-5)-(D-1)x_{1}\right)+(D+1)g^{2}a_{1}^{2}x_{1}(1-x_{1})\right]\nonumber\\
&&\ \ \left.+\frac{\epsilon^{4}g^{2}a_{1}^{2}(1+x_{1})^{2}}{4(1-x_{1})}
\left[(D-3)-(D+1)x_{1}\right]\right\}\frac{d}{dx_{1}}\nonumber\\
&&+\left\{\frac{a_{1}^{2}\omega^{2}(1+x_{1})(1-\epsilon^{2})^{2}}
{4\left[2-g^{2}a_{1}^{2}\left((1-x_{1})+\epsilon^{2}(1+x_{1})\right)\right]}\right.\nonumber\\
&&\ \ +\frac{\left[m_{1}^{2}(1-x_{1})^{2}+m_{2}^{2}\epsilon^{2}(1+x_{1})^{2}\right]
\left[2-g^{2}a_{1}^{2}\left((1-x_{1})+\epsilon^{2}(1+x_{1})\right)\right]}{4(1-x_{1})^{2}(1+x_{1})}\nonumber\\
&&\ \ \left.-\frac{\epsilon^{2}j(j+D-7)}{(1-x_{1})\left[(1-x_{1})+\epsilon^{2}(1+x_{1})\right]}
-\frac{B_{1}\epsilon^{2}(1+x_{1})}{4(1-x_{1})}+\frac{B_{2}}{2(1-x_{1})}\right\},
\end{eqnarray}
and
\begin{eqnarray}
{\cal O}_{2}&=&-\frac{1}{4}(1-x_{2}^{2})\left[2-\epsilon^{2}(1-x_{2})\right]
\left[2-g^{2}a_{1}^{2}\epsilon^{2}(1-x_{2})\right]\frac{d^{2}}{dx_{2}^{2}}\nonumber\\
&&+\left\{\frac{1}{2}\left[(D-7)+(D-3)x_{2}\right]-\frac{\epsilon^{2}}{4}(1+g^{2}a_{1}^{2})(1-x_{2})
\left[(D-5)+(D-1)x_{2}\right]\right.\nonumber\\
&&\ \ \left.+\frac{\epsilon^4}{8}g^{2}a_{1}^{2}(1-x_{2})^{2}\left[(D-3)+(D+1)x_{2}\right]\right\}\frac{d}{dx_{2}}\nonumber\\
&&\left\{\frac{a_{1}^{2}\omega^{2}\epsilon^{2}(1+x_{2})\left[2-\epsilon^{2}(1-x_{2})\right]}
{8\left[2-g^{2}a_{1}^{2}\epsilon^{2}(1-x_{2})\right]}\right.\nonumber\\
&&\ \ +\frac{\left[2-g^{2}a_{1}^{2}\epsilon^{2}(1-x_{2})\right]\left[m_{1}^{2}\epsilon^{2}(1+x_{2})^{2}
+m_{2}^{2}\left(2-\epsilon^{2}(1-x_{2})\right)^{2}\right]}{8(1+x_{2})\left[2-\epsilon^{2}(1-x_{2})\right]}\nonumber\\
&&\ \ \left.+\frac{j(j+D-7)}{2(1-x_{2})}+\frac{B_{1}\epsilon^{2}(1-x_{2})}{8}\right\},
\end{eqnarray}
where
\beq
{\cal O}_{i}R_{i}={B_{i}\over 4} R_{i} ~.
\eeq
We first expand the operators with respect to $\epsilon$. For each equation $i=1,2$ we have:
\begin{eqnarray}
&&\left({\cal O}_{i0}+{\cal O}_{i2}+{\cal O}_{i4}+{\cal O}_{i6}+\cdots\right)\left(R_{i0}+R_{i2}+R_{i4}+R_{i6}+\cdots\right)\nonumber\\
&=&\frac{1}{4}\left(B_{i0}+B_{i2}+B_{i4}+B_{i6}+\cdots\right)\left(R_{i0}+R_{i2}+R_{i4}+R_{i6}+\cdots\right)~,
\label{doublyexp}
\end{eqnarray}
where the next subscript after the $i$ refers to the power of $\epsilon = a_2/a_1$ contained by those terms. Before going on we shall also need to look at the implications of the normalization condition to higher order terms. We first recall that the normalization conditions\footnote{See subsection (\ref{subsec:AIMrotation}).} are given by:
\begin{eqnarray}
  \frac{1}{4}\int^{1}_{-1}(1-x_1)^{\frac{D-5}{2}}R_1^2&=&1\;,\\
  \frac{1}{4}\int^{1}_{-1}(1-x_2)^{\frac{D-7}{2}}R_2^2&=&1\;.
\end{eqnarray}
For convenience we define $\tilde{R}_i=\sqrt{w_i}R_i/2$. Then schematically we have: 
\begin{eqnarray}
  &&\int_{-1}^{1}dx_i\left(\tilde{R}_{i0}+\tilde{R}_{i2}+\tilde{R}_{i4}+\tilde{R}_{i6}+\cdots\right)\left(\tilde{R}_{i0}+\tilde{R}_{i2}+\tilde{R}_{i4}+\tilde{R}_{i6}+\cdots\right)=1\;,\label{orthoexp}
\\
  &\Rightarrow&\int^1_{-1}dx_i\tilde{R}_{i0}^2=1\;; \quad \int_{-1}^{1}dx_i\ \tilde{R}_{i0}\tilde{R}_{i2}=0\;;\quad \int_{-1}^{1}dx_1\ \tilde{R}_{i0}\tilde{R}_{i4}=-\frac{1}{2}\int_{-1}^{1}dx_1\ \tilde{R}_{i2}^2\;;\quad \int_{-1}^{1}dx_i\ \tilde{R}_{i0}\tilde{R}_{i6}=-\int_{-1}^{1}dx_i\ \tilde{R}_{i2}\tilde{R}_{i4}\;.\nonumber\end{eqnarray}

\subsection{Zeroth order in epsilon $\mathcal{O}(\epsilon^0)$ }
To the zeroth order in $\epsilon$, the eigenvalue equations are 
\beq
{\cal O}_{i0}R_{i0}={B_{i0}\over 4} R_{i0}, 
\label{zeroth}
\eeq
where,
\begin{eqnarray}
{\cal O}_{10}&=&
-\frac{(1-x_{1}^{2})}{2}\left[2-g^{2}a_{1}^{2}(1-x_{1})\right]\frac{d^{2}}{dx_{1}^{2}}
+\left[\frac{1}{2}\left((D-5)+(D-1)x_{1}\right)
-\frac{1}{4}g^{2}a_{1}^{2}(1-x_{1})\left((D-3)+(D+1)x_{1}\right)\right]
\frac{d}{dx_{1}}\nonumber\\
&&\ \ +\left\{\frac{a_{1}^{2}\omega^{2}(1+x_{1})}{4[2-g^{2}a_{1}^{2}(1-x_{1})]}
+\frac{m_1^2 [2-g^{2}a_{1}^{2}(1-x_{1})]}{4(1+x_{1})}+\frac{B_{20}}{2(1-x_{1})}\right\},\\
{\cal O}_{20}
&=&-(1-x_{2}^{2})\frac{d^{2}}{dx_{2}^{2}}+\frac{1}{2} (D-7 + (D- 3)x_2)\frac{d}{dx_{2}}
+\frac{m_2^2}{2(1+x_{2})}+ {j (j+D-7)\over2(1-x_2)}\equiv {\cal O}_{200}~.
\label{O20}
\end{eqnarray}

We first consider the ${\cal O}_{20}$ equation. We note that the ${\cal O}_{20}$ operator does not involve powers of $a_{1}$ at $\mathcal{O}(\epsilon^0)$ order, therefore we can write ${\cal O}_{20}={\cal O}_{200}+{\cal O}_{202}+\dots={\cal O}_{200}$ where the third subscript refers to the power of $a_1$ in those terms. To fix notation we also redefine the solution $R_{20}=R_{200}$ and the eigenvalue $B_{20}=B_{200}$.  
\par One then observes that the corresponding eigenvalue equation 
\beq
{\cal O}_{200}R_{200}={B_{200}\over 4}R_{200}\;,
\eeq
is exactly solvable in terms of Jacobi polynomials. The solution is given by
\begin{eqnarray}
R_{200}(n_2)&=&c_{2n_2m_2j}(1-x_{2})^{j/2}(1+x_{2})^{|m_2|/2}P^{(j+{D-7\over 2},|m_2|)}_{n_{2}}(x_{2})\;,\nn\\
B_{200}(n_2)&=& (2 n_2+|m_2|+ j) (2n_2+|m_2|+ j +D-5 )\;,
\label{B200}
\end{eqnarray}
where $n_{2}=0,1,2,\dots$. The normalization condition:
\begin{eqnarray}
\frac{1}{4}\int_{-1}^{1}dx_{2}(1-x_2)^{(D - 7 )\over 2} R_{100}(n_{1})R_{100}(n'_{1})=\delta_{n_{1}n'_{1}}\,,
\end{eqnarray} 
is satisfied if \footnote{This can be found using the orthonormality of the Jacobi polynomials.} 
\beq
c_{2n_2m_2j}=\left\{ \frac{(2 j+2
   |m_2|+4 n_2+D-5) \Gamma(n_2+1) \Gamma
   \left(j+|m_2|+n_2+\frac{D-5}{2}\right)}{2^{\frac{1}{2} (2j+2 |m_2|+D-7)}  \Gamma
   (|m_2|+n_2+1) \Gamma \left(j+n_2+\frac{D-5}{2}\right)}\right\}^{1/2}.
\eeq
For the properties of Jacobi polynomials see  \cite{Gradshteyn:2000}.

\par Next we work on the ${\cal O}_{10}$ equation. Following the same method we have used for $\epsilon$ we expand the eigenvalue equation in powers of $a_{1}$. The operator at zeroth order is
\begin{eqnarray}
{\cal O}_{100}&=&-(1-x_{1}^{2})\frac{d^{2}}{dx_{1}^{2}}+\frac{1}{2}(D-5 + (D-1) x_1)\frac{d}{dx_{1}}+ {B_{200}\over 2 (1 - x_1)} + { m_1^2\over 2(1 +  x_1)}~,
\label{O100}
\end{eqnarray}
where, unlike ${\cal O}_{200}$,  the ${\cal O}_{100}$ operator is coupled to ${\cal O}_{200}$ through $B_{200}$.

\par The zeroth order eigenvalue equation ${\cal O}_{100}R_{100}=\frac 1 4 B_{100}R_{100}$ can again be solved and the solutions and eigenvalues are found to be:
\begin{eqnarray}
R_{100}(n_1)&=&c_{1n_1m_1}(1-x_1)^{(2\kappa + 5 - D )\over 4} (1+x_1)^{|m_1|\over 2}P_{n_{1}}^{( \kappa ,|m_1|)}(x_{1})\,,
\nn\\
B_{100}(n_1)
 &=&( 2(n_1 + n_2)+ |m_1| + |m_2|+j )(2(n_1 + n_2)+ |m_1| + |m_2|+j + D-3)\;, 
 \label{B100}
\end{eqnarray}
where $n_{1}=0,1,2,\dots$, and for clarity we have defined
\beq
\kappa=\sqrt{ B_{200} + \Big({D-5\over2}\Big)^2}= 2n_2+|m_2|+  j + \frac 1 2 (D-5) \;.
\eeq
The orthonormality condition 
\begin{eqnarray}\label{eqn:orthonR1}
\frac{1}{4}\int_{-1}^{1}dx_{1}(1-x_1)^{(D - 5 )\over 2} R_{100}(n_{1})R_{100}(n'_{1})=\delta_{n_{1}n'_{1}}\,,
\end{eqnarray}
is satisfied with the normalisation 
\beq
c_{1n_1m_1}=\left\{
\frac{ (2 n_1+\kappa+|m_1|+1) \Gamma (n_1+1) \Gamma
   \left(n_1+\kappa+|m_1|+1\right)}{2^{\kappa+|m_1|-1}\Gamma \left(n_1+\kappa+1\right) \Gamma (n_1+|m_1|+1)}\right\}^{1/2}.
\label{zeronorm}
\eeq

\par We now describe how given both the zeroth order eigenvalues and eigenfunctions we can go to higher order in the perturbative series.

\subsubsection*{Second order in $a_1^2$, $\mathcal{O}{(\epsilon^0,a_1^2)}$}

At next order in the perturbative expansion for ${\cal O}_1$ we find:
\beq\label{eqn:O102evalueExp}
{\cal O}_{102}R_{100}+{\cal O}_{100}R_{102}=\frac{1}{4}\left(B_{102}R_{100}+B_{100}R_{102}\right).
\eeq
\par Where the ${\cal O}_{102}$ operator is found to be:
\begin{eqnarray}\label{eqn:O102}
{\cal O}_{102}=\frac{1}{2}a_1^2g^{2}(1-x_{1})(1-x_{1}^{2})\frac{d^{2}}{dx_{1}^{2}}-\frac{1}{4}a_1^2g^{2}(1-x_{1})(D-3 + (D+1)x_1)\frac{d}{dx_{1}}
+\frac 1 8 a_1^2\omega^2  (1+x_1)
-\frac 1 4 a_1^2 g^{2}m_1^2{1-x_1\over 1+x_1}\nn\;.
\end{eqnarray}

\par We also need to look at the effect of the normalization condition (\ref{eqn:orthonR1}) to higher order terms. We again find a series of conditions analogous to (\ref{orthoexp}), i.e.,  
\begin{eqnarray}
  \int^1_{-1}dx_1\tilde{R}_{100}^2=1\;; \quad \int_{-1}^{1}dx_1\ \tilde{R}_{100}\tilde{R}_{102}=0\quad\cdots.\end{eqnarray}
  Using these relations and the hermiticity of the operators we are able to find $B_{102}$ by contracting equation (\ref{eqn:O102evalueExp}) with $R_{100}$:
\begin{eqnarray}
B_{102}=\int_{-1}^{1}dx_{1}(1-x_1)^{(D - 5 )\over 2}  R_{100}(n_{1}){\cal O}_{102}R_{100}(n'_{1})~.
\end{eqnarray}
Using the equation ${\cal O}_{100}R_{100}=\frac{1}{4}B_{100}R_{100}$ to remove the second derivative we obtain:  
\begin{eqnarray}
{\cal O}_{102} R_{100} =\Big[-\frac 1 2 a_1^2g^2  (1-x_1^2) \frac{d}{dx_{1}}+ \frac 1 4 a_1^2g^2 B_{200}+\frac 1 8 a_1^2\omega^2 (1+x_1)-\frac 1 8 B_{100} a_1^2g^2 (1-x_1)
\Big] R_{100}~.
\label{O102}
\end{eqnarray}
After applying various identities for Jacobi polynomials \cite{Gradshteyn:2000} to remove the derivative and $x_1$ dependence in equation (\ref{O100}) we obtain:
\begin{eqnarray}
B_{102}&=&\frac{a_1^2}{2} (\omega^2 +  g^2 B_{100}+4 g^2 n_1 + g^2 (2 \kappa + 5 - D) + 2 g^2 |m_1| ) \left(\frac{\left(m_1^2-\kappa ^2\right)}{(2 n_1+|m_1|+\kappa )(2 n_1+|m_1|+\kappa +2)}\right)   \nn\\
  &&+
  \frac{a_1^2}{2} (\omega^2 + 2 g^2 B_{200} - g^2 B_{100} + 
   g^2 (2 \kappa + 5 - D) - 2 g^2 |m_1|) - 2 a_1^2g^2 n_1\frac{(\kappa-|m_1|)}{2n_1+|m_1|+\kappa }~.
   \label{B102}
\end{eqnarray}

\par Note that in the single rotation limit ($a_2\to 0$) with $g=0$ we find agreement with the result obtained in \cite{Berti:2005gp} using inverted continued fractions. The result does not, however, agree with that for $g\neq 0$ using inverted fractions in powers of $c_1=a_1\omega$ and $\alpha_1=a_1^2 g^2$ \cite{Cho:2009wf} because we used an expansion in different parameters: $\epsilon = a_2/a_1$ and $a_1$.\footnote{We have verified that perturbation theory (using orthogonal polynomials) for small $c_1,c_2$ and $\alpha_1,\alpha_2$ does indeed give the correct second order answer when $c_2,\alpha_2\to 0$ for $D=5$ \cite{Cho:inPROG, Aliev:2008yk} (cf. \cite{Cho:2009wf} using inverted continued fractions).} 

\par  In theory we could continue to go to higher order in $a_1^2$, however, we stop at this order, to point out an issue that arises in the general case for the second order in $\epsilon$ terms.   

\subsection{Second order in epsilon $\mathcal{O}(\epsilon^2)$}

\par We now consider the next order in $\epsilon$, firstly for ${\cal O}_{2}$ we have the operator equation,
\beq
{\cal O}_{22}R_{20}+{\cal O}_{20}R_{22}=\frac{1}{4}\left(B_{20}R_{22}+B_{22}R_{20}\right)~.
\eeq
After using the orthonormality of the doubly perturbed eigenfunctions up to second order in $a_1^2$, the hermiticity of the operators and also the fact that many of the terms are simply zero in the ${\cal O}_{2}$ case, we find again that:
\beq
B_{22i}=\int_{-1}^{1}dx_2 (1-x_2)^{D-7\over 2}\ \!R_{200}{\cal O}_{22i}R_{200}~,
\eeq
where $i=0,2$. 

\par The $\epsilon^2$ operator takes the following form:
\bea
{\cal O}_{22}
&=&\frac{\epsilon^2}{2} (1+x_2) (1-x_2)^2 \left(1+a_1^2 g^2\right)\frac{d^{2}}{dx_{2}^{2}}
   -\frac{\epsilon^2}{4} (1-x_2) \left(1+a_1^2 g^2\right) (D-5+(D-1)x_2) \frac{d}{dx_{2}}
   \nn\\
&&+ \frac{\epsilon^2}{8}  (1+x_2) (a_1^2\omega^2+m_1^2)+\frac{\epsilon^2}{8} B_{10}
   (1-x_2)-\epsilon^2\left(1+a_1^2 g^2\right)\frac{m_2^2(1-x_2) }{4(1+x_2)}\;.
   \nn
   \eea
Thus, working up to second order in $a_1^2$, we have:
\bea
{\cal O}_{220}R_{200} &=& \Big[ -\frac{\epsilon^2}{2} (1-x_2^2) \frac{d}{dx_{2}}+\frac{\epsilon^2}{8} m_1^2 (1+x_2) +\frac{\epsilon^2}{8} B_{100} (1-x_2)+\frac{\epsilon^2}{4}j(j+D-7)-\frac{ \epsilon^2}{ 8}  B_{200}(1-x_2)\Big] R_{200}\nn
 \\
{\cal O}_{222} R_{200} &=&\Big[ -\frac{1}{2} (1-x_2^2) \epsilon^2 a_1^2g^2 \frac{d}{dx_{2}}+\frac{ \epsilon^2 a_1^2}{8} \omega^2(1+x_2)+\frac{ \epsilon^2}{8} B_{102} (1-x_2) +\frac{1}{4}  \epsilon^2 a_1^2g^2j(j+D-7) \nonumber \\
&&-\frac 1 8 \epsilon^2 a_1^2g^2 B_{200}(1-x_2)\Big] R_{200}\;,
\eea
where in the above steps we used ${\cal O}_{200}R_{200}$, cf. equation (\ref{O20}), to remove second derivatives. 
Again using the functional properties of the Jacobi polynomials \cite{Gradshteyn:2000} we find:
\bea
B_{220}&=&\frac{\epsilon^2}{ 2} \left(m_1^2 - B_{100} + B_{200} + 4 n_2 + 2 j + 2 |m_2|\right)\frac{\left(m_2^2-\alpha ^2\right)}{(2n_2 +|m_2|+\alpha )(2 n_2 +|m_2|+\alpha +2) } 
\nn\\
&&+
\frac{\epsilon^2}{2} \left(m_1^2-B_{200}+B_{100}\right) +\epsilon^2j (D+j-7) -2\epsilon^2 n_2 \frac{ (\alpha -|m_2|)}{2n_2+|m_2|+\alpha }+\epsilon^2(j-|m_2|)~,
      \label{B220}
   \\
   B_{222}&=&\frac{\epsilon^2}{2} \left(a_1^2\omega^2 - B_{102} + a_1^2g^2 B_{200} + 4 a_1^2g^2 n_2 + 2 a_1^2g^2 j + 2a_1^2g^2  |m_2|\right)\frac{\left(m_2^2-\alpha ^2\right)}{(2n_2 +|m_2|+\alpha )
   (2 n_2 +|m_2|+\alpha +2) } 
   \nn\\
   &&+\frac{\epsilon^2}{2} \left(a_1^2\omega^2 + B_{102} - a_1^2g^2 B_{200}\right) 
   + \epsilon^2a_1^2g^2 j (D+j-7) -2\epsilon^2a_1^2g^2 n_2 \frac{(\alpha -|m_2|)}{2n_2+|m_2|+\alpha }
   +\epsilon^2a_1^2g^2(j-|m_2|)~, \nonumber \\ &&
 \label{B222}
 \eea
where we defined $\alpha = j + (D - 7)/2$. 

\subsubsection*{Second order in ${\cal O}_{1}$}

\par It is in the $\mathcal{O}(\epsilon^2)$ order of the ${\cal O}_1$ operator that our method runs into some difficulty in the general case. For the ${\cal O}_{1}$ operator at $\mathcal{O}(\epsilon^2,a_1^0)$ order we find:
\beq
{\cal O}_{120}R_{100}+{\cal O}_{100}R_{120}=\frac{1}{4}\left(B_{100}R_{120}+B_{120}R_{100}\right)~,
\label{O12pert}
\eeq
where after performing manipulations similar to those in the previous sections we find:
\bea
B_{120}&=&\int_{-1}^{1}dx_1 (1-x_1)^{(D - 5 )\over 2}\!R_{100}{\cal O}_{120}R_{100}~.
\label{B120tildenorm}
\eea
As such, an expansion of ${\cal O}_{12}$ in powers of $a_1^2$ leads to
\bea
{\cal O}_{120}   &=&-\epsilon^2\left(1+x_1\right)^2{d^2\over dx_1^2}
-\frac{1}{2} \epsilon^2(D-5 - (D-1)x_1){1+x_1\over 1-x_1}{d\over dx_1}
+\frac{\epsilon^2m_2^2}{2}\frac{(1+x_1)}{(1-x_1)^2}-\epsilon^2\frac{j (j+D-7)}{ (1-x_1)^2}\nonumber \\
&&+{B_{220}\over2(1-x_1)} -\epsilon^2{1 + x_1 \over 1 - x_1} {B_{100}\over 4}.
\nn\\
\eea
Now we can use ${\cal O}_{100} R_{100}$, see equation (\ref{O100}), to remove second derivative terms, and we obtain:
 \bea
 B_{120}&=&\int_{-1}^{1}dx_1 (1-x_1)^{(D - 5 )\over 2}\!R_{100}\Big[
 -\epsilon^2(D-5){(1+x_1)\over  (1 - x_1)} \frac d {dx_1} +\epsilon^2\frac{(m_2^2-B_{200})(1+x_1)-2j (j+D-7)}{2(1-x_1)^2} \nonumber \\
 && \hspace{5cm}
 +{B_{220}-\epsilon^2 m_1^2\over 2 (1 -x_1)} 
 \Big]R_{100}~.
 \nn\\
 \label{B120}
 \eea
\par Unfortunately, the terms with factors of $(1-x)$ in the denominator do not appear to allow any simplification via standard Jacobi identities \cite{Gradshteyn:2000}. Thus, in the general case it does not seem possible to find the solution to $B_{120}$ in closed algebraic form. In principle one could still numerically integrate these expressions, however, as explained in section (\ref{numeric}), when calculating the QNM's it is more advantageous to perform a full numerical calculation than to take this route.

\par Nevertheless, we have found that the offending terms vanish for the special choice of parameters $D=6$, $j=0$, $m_{1}=m_{2}=1$ and $n_{1}=n_{2}=0$, and in this case, $B_{120}=0$.
\subsection{Sixth order in the special case: $D=6$, $j=0$, $m_{1}=m_{2}=1$ and $n_{1}=n_{2}=0$}
We also found that this special case allowed us to go to higher orders without encountering the above mentioned issue. To go to higher orders equation's like (\ref{B120tildenorm}) often can not be written simply in terms of the zeroth order eigenfunctions. In such cases we had to decompose the higher order functions in terms of linear superpositions of zeroth order ones. We will give a more detailed explanation of this method in \cite{Cho:inPROG}. Here we just list the result:
\begin{eqnarray}
  B_{1}&=&\left[10+\frac{2}{9}\left(2\omega^{2}-17g^{2}\right)a_{1}^{2}-\frac{20}{8019}(\omega^{4}-53g^{2}\omega^{2}+196g^{4})a_{1}^{4}\right.\nonumber\\
  &&\ \ \ \ \ \ \ \ \ \ \left.+\frac{20}{8444007}(2\omega^{6}-645g^{2}\omega^{4}+28959g^{4}\omega^{2}-105644g^{6})a_{1}^{6}\right]\nonumber\\
  &&+\epsilon^{2}\left[\frac{2}{9}(2\omega^{2}-17g^{2})a_{1}^{2}+\frac{32}{8019}(\omega^{4}-53g^{2}\omega^{2}+196g^{4})a_{1}^{4}
  -\frac{32}{2814669}(\omega^{6}-147g^{2}\omega^{4}+5178g^{4}\omega^{2}-18424g^{6})a_{1}^{6}\right]\nonumber\\
  &&+\epsilon^{4}\left[-\frac{20}{8019}(\omega^{4}-53g^{2}\omega^{2}+196g^{4})a_{1}^{4}
  -\frac{32}{2814669}(\omega^{6}-147g^{2}\omega^{4}+5178g^{4}\omega^{2}-18424g^{6})a_{1}^{6}\right]\nonumber\\
  &&+\epsilon^{6}\left[\frac{20}{8444007}(2\omega^{6}-645g^{2}\omega^{4}+28959g^{4}\omega^{2}-105644g^{6})a_{1}^{6}\right]+\cdots,\\
  B_{2}&=&2+\epsilon^{2}\left[2+\frac{2}{9}(4\omega^{2}-7g^{2})a_{1}^{2}-\frac{4}{8019}(\omega^{4}-53g^{2}\omega^{2}+196g^{4})a_{1}^{4}\right.\nonumber\\
  &&\ \ \ \ \ \ \ \ \ \ \ \ \ \left.+\frac{4}{8444007}(2\omega^{6}-645g^{2}\omega^{4}+28959g^{4}\omega^{2}-105644g^{6})a_{1}^{6}\right]\nonumber\\
  &&+\epsilon^{4}\left[-\frac{4}{8019}(\omega^{4}-53g^{2}\omega^{2}+196g^{4})a_{1}^{4}
  -\frac{32}{8444007}(4\omega^{2}-25g^{2})(\omega^{4}-53g^{2}\omega^{2}+196g^{4})a_{1}^{6}\right]\nonumber\\
  &&+\epsilon^{6}\left[\frac{4}{8444007}(2\omega^{6}-645g^{2}\omega^{4}+28959g^{4}\omega^{2}-105644g^{6})a_{1}^{6}\right]+\cdots.
\end{eqnarray}

\par These results are compared to the AIM in figures \ref{fig:epsilon1} and \ref{fig:largeomega}. In figure \ref{fig:epsilon1} we see that $\sim 80$ iterations are required to be as good as the perturbative value, where a similar plot was found for $b_2$. In figure \ref{fig:largeomega} we see that the perturbative eigenvalues break down for $\omega$ larger than $\sim 5$.

\bibliography{Kerr}{}

\end{document}